\documentclass[manuscript]{acmart}

\AtBeginDocument{%
  }

\setcopyright{acmlicensed}
\copyrightyear{2026}
\acmYear{2026}
\acmDOI{XXXXXXX.XXXXXXX}
\acmConference[CHI'26]{CHI conference on Human Factors in Computing Systems}{June 03--05,
  2026}{Barcelona}
  
\acmISBN{978-1-4503-XXXX-X/2018/06}

\usepackage[T1]{fontenc}

\usepackage{graphicx} 
\usepackage{color,soul}
\usepackage{paralist}
\usepackage{wrapfig}
\usepackage{tabularx}
\usepackage{multirow}
\usepackage{booktabs}
\usepackage{subcaption}
\usepackage{xcolor}
\usepackage{colortbl}
\usepackage{fontawesome5}
\usepackage{balance}
\usepackage{framed}
\usepackage{longtable}
\usepackage{url}
\usepackage{booktabs}
\usepackage{amsmath}
\usepackage{xspace}
\usepackage{framed}        % for \MakeFramed, \FrameCommand
\usepackage{changepage}    % for adjustwidth

\usepackage[most]{tcolorbox}

\everypar{\looseness=-1}
\newcommand\gencite[3]{\cite[#3]{#1}}
\newcommand\genbibitem[3]{\bibitem{#1} #3}

\newcommand{\surprise}{\faSurprise} 
\newcommand{\concrete}{\faAddressCard} 
\newcommand{\severe}{\faSkullCrossbones} 
\newcommand{\relevant}{\faBullseye} 
\newcommand{\diverse}{\faPalette} 
\newcommand{\champion}{\faUserPlus}
\newcommand{\resistor}{\faUserMinus} 
\newcommand{\indifferent}{\faUser} 
\newcommand{\follower}{\faUserTie}
\newcommand{\learner}{\faUserGraduate}

\definecolor{sticky}{HTML}{00798c}  
\definecolor{baseline}{HTML}{edae49}

\newcommand{\stickyP}[1]{\textcolor{sticky}{\textbf{\textit{#1}}}}
\newcommand{\baselineP}[1]{\textcolor{baseline}{\textbf{\textit{#1}}}}

\newcommand{\POne}{\stickyP{P1}\xspace} 
\newcommand{\PTwo}{\stickyP{P2}\xspace} 
\newcommand{\PThree}{\baselineP{P3}\xspace} 
\newcommand{\PFour}{\stickyP{P4}\xspace} 
\newcommand{\PFive}{\stickyP{P5}\xspace} 
\newcommand{\PSix}{\baselineP{P6}\xspace} 
\newcommand{\PSeven}{\baselineP{P7}\xspace} 
\newcommand{\PEight}{\stickyP{P8}\xspace} 
\newcommand{\PNine}{\stickyP{P9}\xspace} 
\newcommand{\PTen}{\baselineP{P10}\xspace} 
\newcommand{\PEleven}{\baselineP{P11}\xspace} 
\newcommand{\PTwelve}{\stickyP{P12}\xspace} 
\newcommand{\PThirteen}{\stickyP{P13}\xspace} 
\newcommand{\PFourteen}{\baselineP{P14}\xspace} 
\newcommand{\PFifteen}{\stickyP{P15}\xspace} 
 
\newcommand{\PSeventeen}{\stickyP{P17}\xspace} 
 
\newcommand{\PNineteen}{\stickyP{P19}\xspace} 
\newcommand{\PTwenty}{\baselineP{P20}\xspace} 
\newcommand{\PTwentyOne}{\stickyP{P21}\xspace} 
\newcommand{\PTwentyTwo}{\baselineP{P22}\xspace} 
\newcommand{\PTwentyThree}{\stickyP{P23}\xspace} 
\newcommand{\PTwentyFour}{\stickyP{P24}\xspace} 
\newcommand{\PTwentyFive}{\baselineP{P25}\xspace} 
\newcommand{\PTwentySix}{\stickyP{P26}\xspace} 
\newcommand{\PTwentySeven}{\baselineP{P27}\xspace} 
\newcommand{\PTwentyEight}{\baselineP{P28}\xspace} 
\newcommand{\PTwentyNine}{\baselineP{P29}\xspace} 
\newcommand{\PThirty}{\stickyP{P30}\xspace} 
\newcommand{\PThirtyOne}{\baselineP{P31}\xspace}

\definecolor{PAblue}{RGB}{0,122,204}%{HTML}{007acc}
\newcommand{\quo}[1]{ 
        \vspace{-0.1cm}
        \def\FrameCommand{%
                \hspace{8pt}%
                {\color{PAblue}\vrule width 2pt}%
                {\color{white}\vrule width 2pt}%
                \colorbox{white}
        }%
        \MakeFramed{\advance\hsize-\width\FrameRestore}%
        \noindent\hspace{-4.55pt}% disable indenting first paragraph
        \begin{adjustwidth}{}{0pt}
                {#1}
                \vspace{-3pt}
        \end{adjustwidth}\endMakeFramed%
}

\newcommand\trajectory[4]{
\begin{tcolorbox}[colback=gray!3!white,
                  colframe=gray!60!black,
                  boxrule=0.6pt,arc=2mm,
                  auto outer arc, left=4pt, right=4pt, width=0.95\linewidth]
\centering
#1 $\;\longrightarrow\;$ #2 $\;\longrightarrow\;$ #3%
\if\relax\detokenize{#4}\relax
\else
  $\;\longrightarrow\;$ #4%
\fi
\end{tcolorbox}
}

\newcommand\figureOverview[0]{
\begin{figure}[!h]
\centering
\includegraphics[width=1\linewidth]{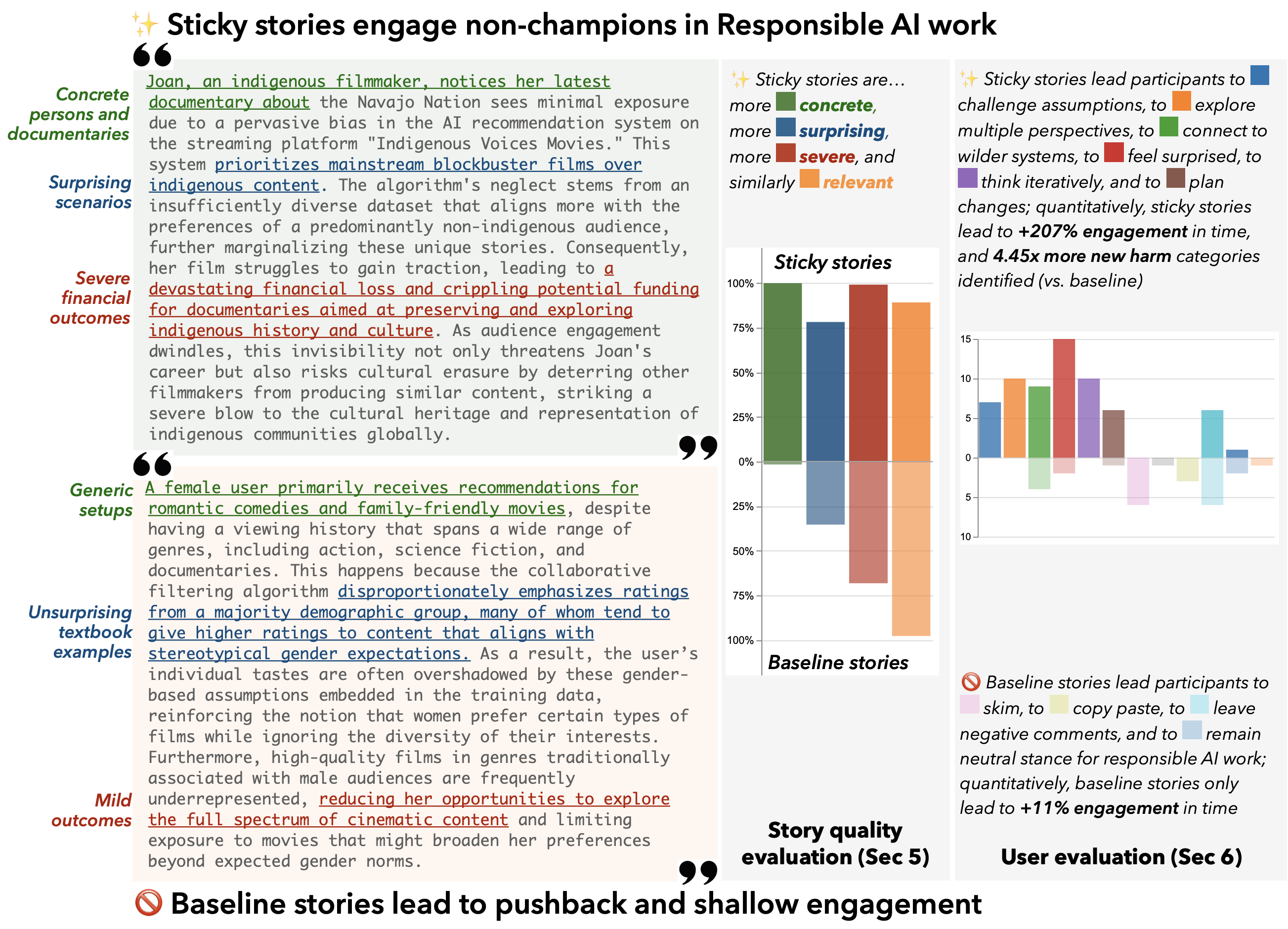}
\caption{Examples of different types of stories: (Bottom) Baseline story, which tends to be generic and straightforward but less memorable, and (Top) Sticky story, which incorporates elements such as surprise and emotional resonance to encourage deeper engagement and recall. We design (Section 4) and validate (Section 5) the sticky stories to be concrete, surprising, and severe. Our user evaluation (Section 6) with RAI practitioners demonstrates that sticky stories can better engage non-champions and lead them to more critical reflections.}
\label{fig:overview}
\end{figure}
}

\newcommand\figureRAI[0]{
\begin{figure}[h!]
\centering
\includegraphics[width=1\linewidth]{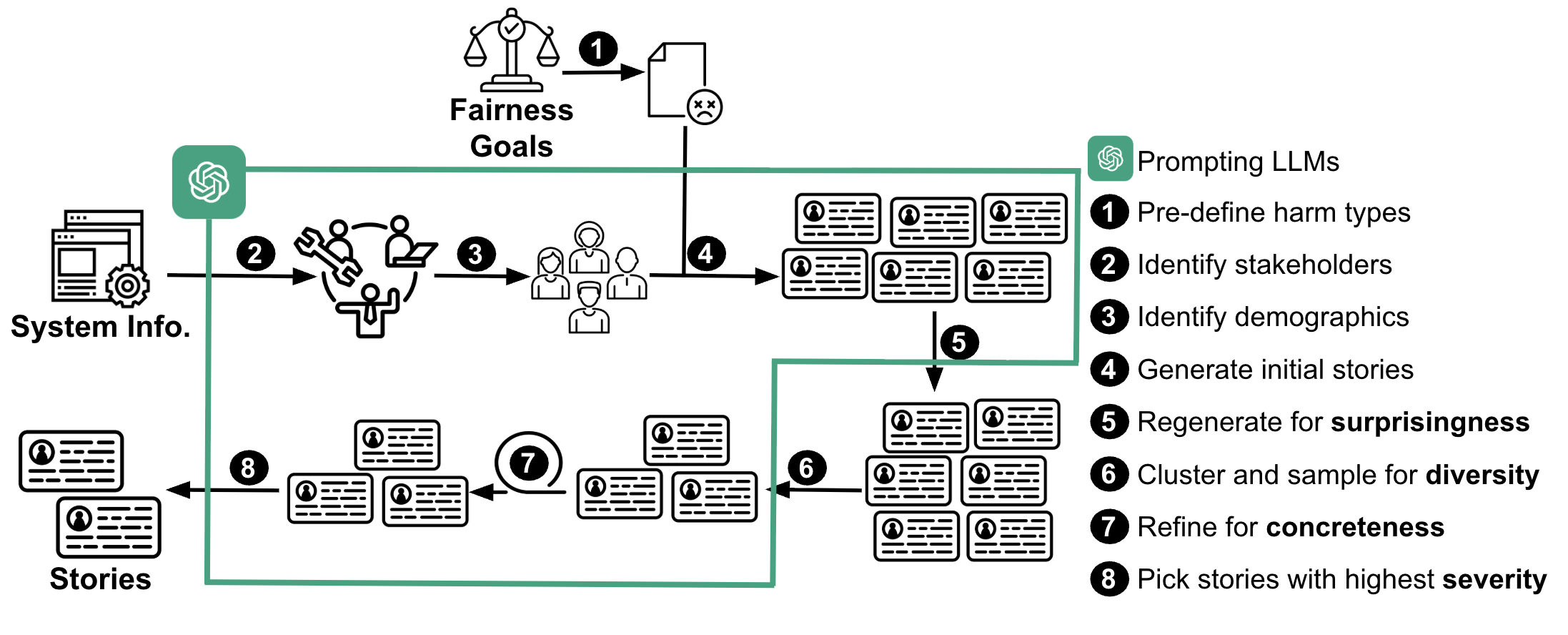}
\caption{Pipeline for \textit{Sticky Story} Generation}
\label{fig:rai}
\end{figure}
}

\newcommand\figureTool[0]{
\begin{figure}[t]
\centering
\includegraphics[width=1\linewidth]{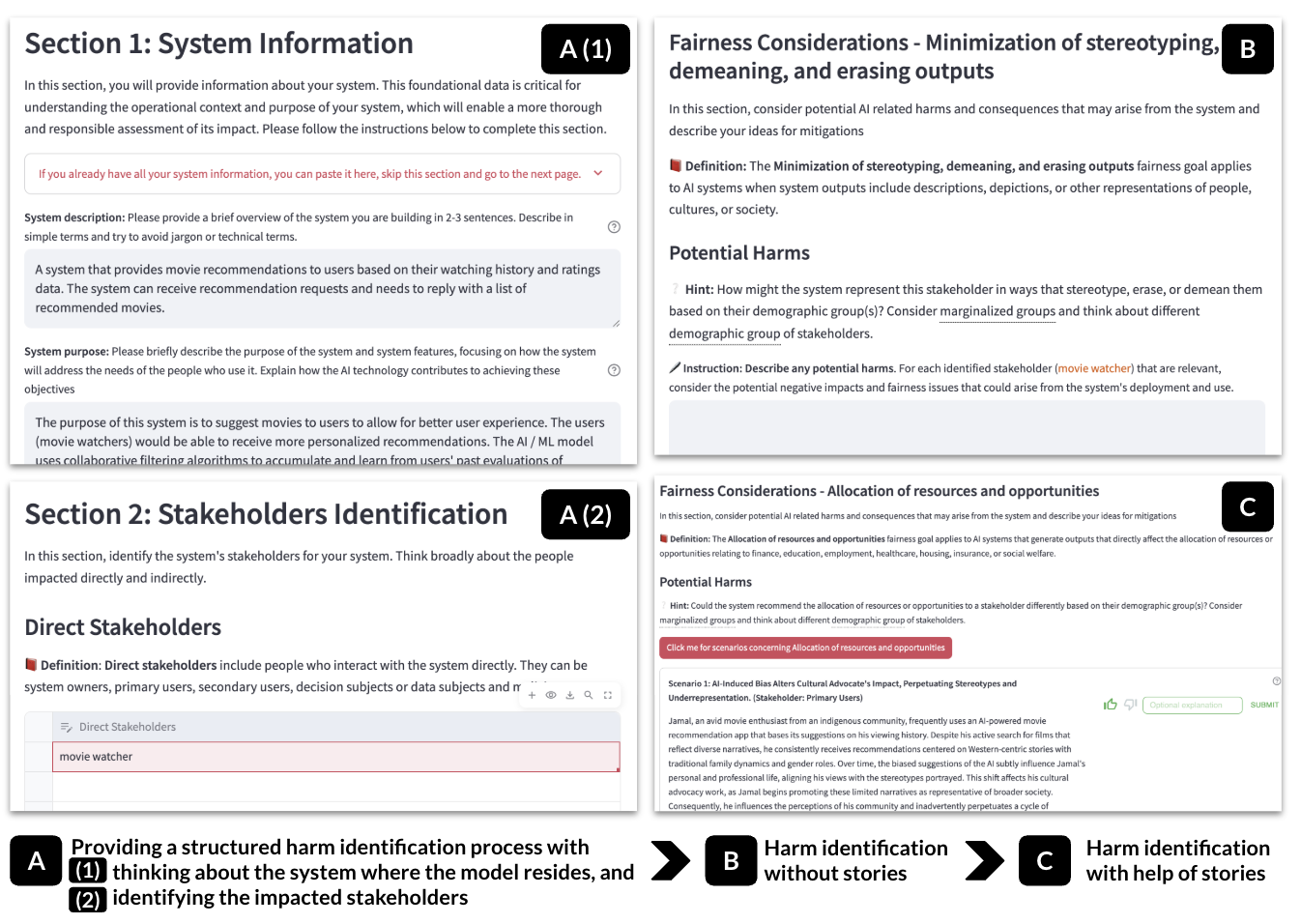}
\caption{Snapshots of the Tool Integrating the Sticky Story Generation Pipeline}
\label{fig:tool}
\end{figure}
}

\newcommand\figureCodes[0]{
\begin{figure}[t]
\centering
\includegraphics[width=1\linewidth]{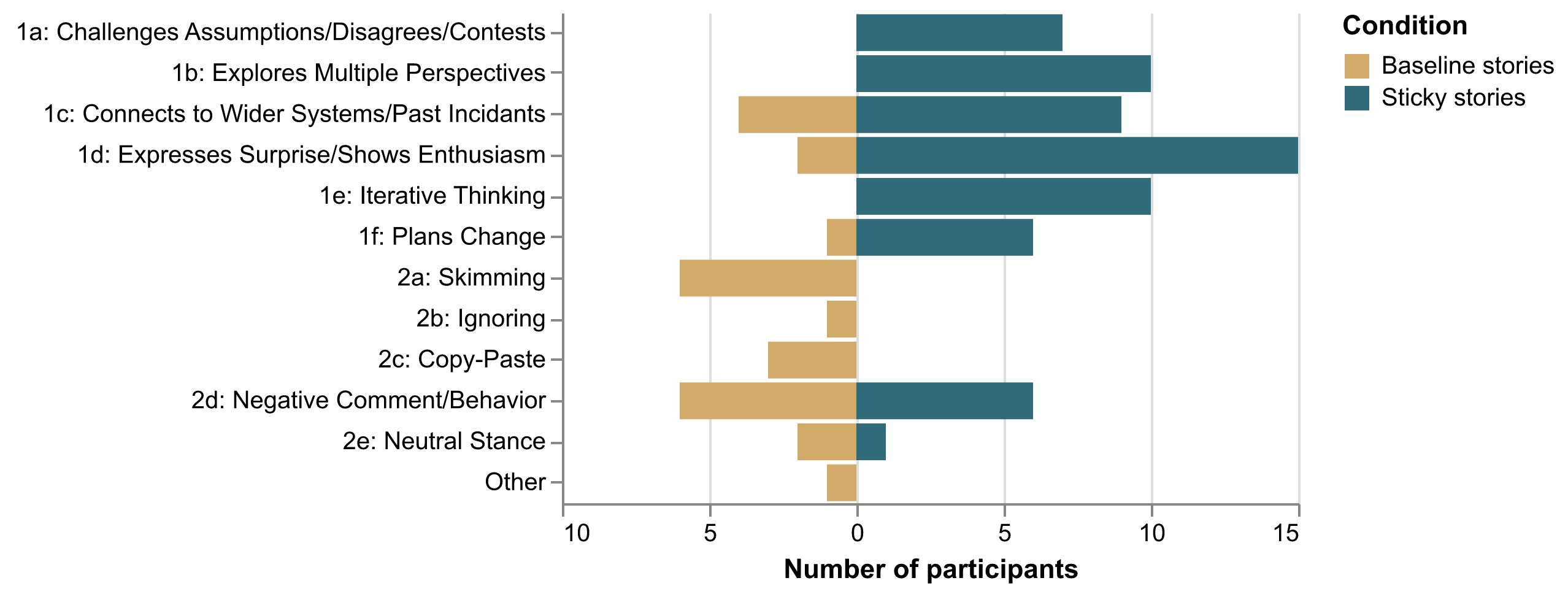}
\caption{Comparison of code frequencies per participant across baseline and sticky story conditions, highlighting increased reflection in the sticky story condition.}
\label{fig:codes}
\end{figure}
}

\newcommand\figureViolin[0]{
\begin{figure}[t]
\centering
\includegraphics[width=1\linewidth]{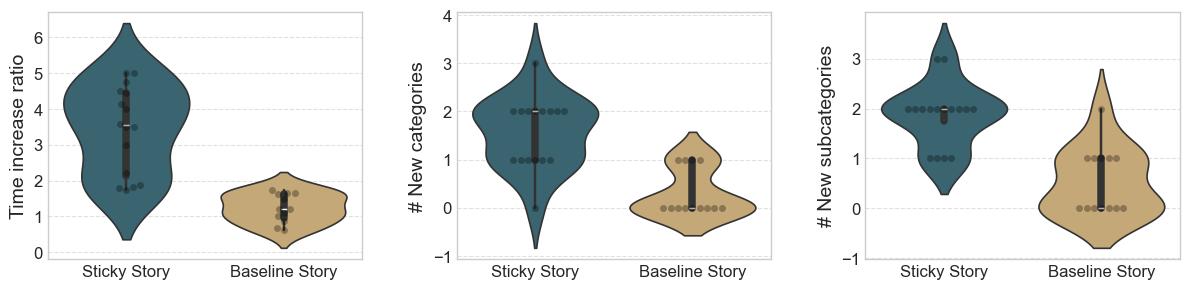}
\caption{We found that participants spent more time and identified more new harms in distinct categories or subcategories, after they read sticky stories.}
\label{fig:violin}
\end{figure}
}

\newcommand\figureMatrices[0]{
\begin{figure}[t]
\centering
\includegraphics[width=1\linewidth]{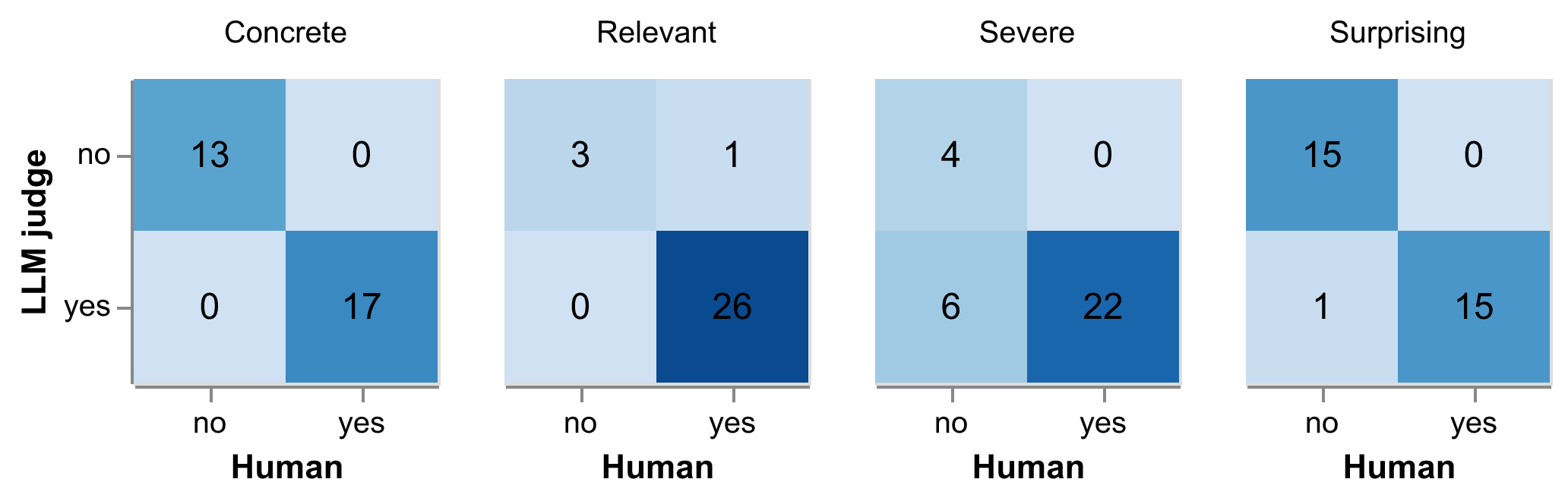}
\caption{Humans agree well with LLM judges on whether a story in concrete, relevant, or surprising. While there is more disagreement on severity, we found they all stemmed from LLM being overly generous in rating baseline stories as severe.}
\label{fig:matrices}
\end{figure}
}

\newcommand\figureHistogram[0]{
\begin{figure}[t]
\centering
\includegraphics[width=1\linewidth]{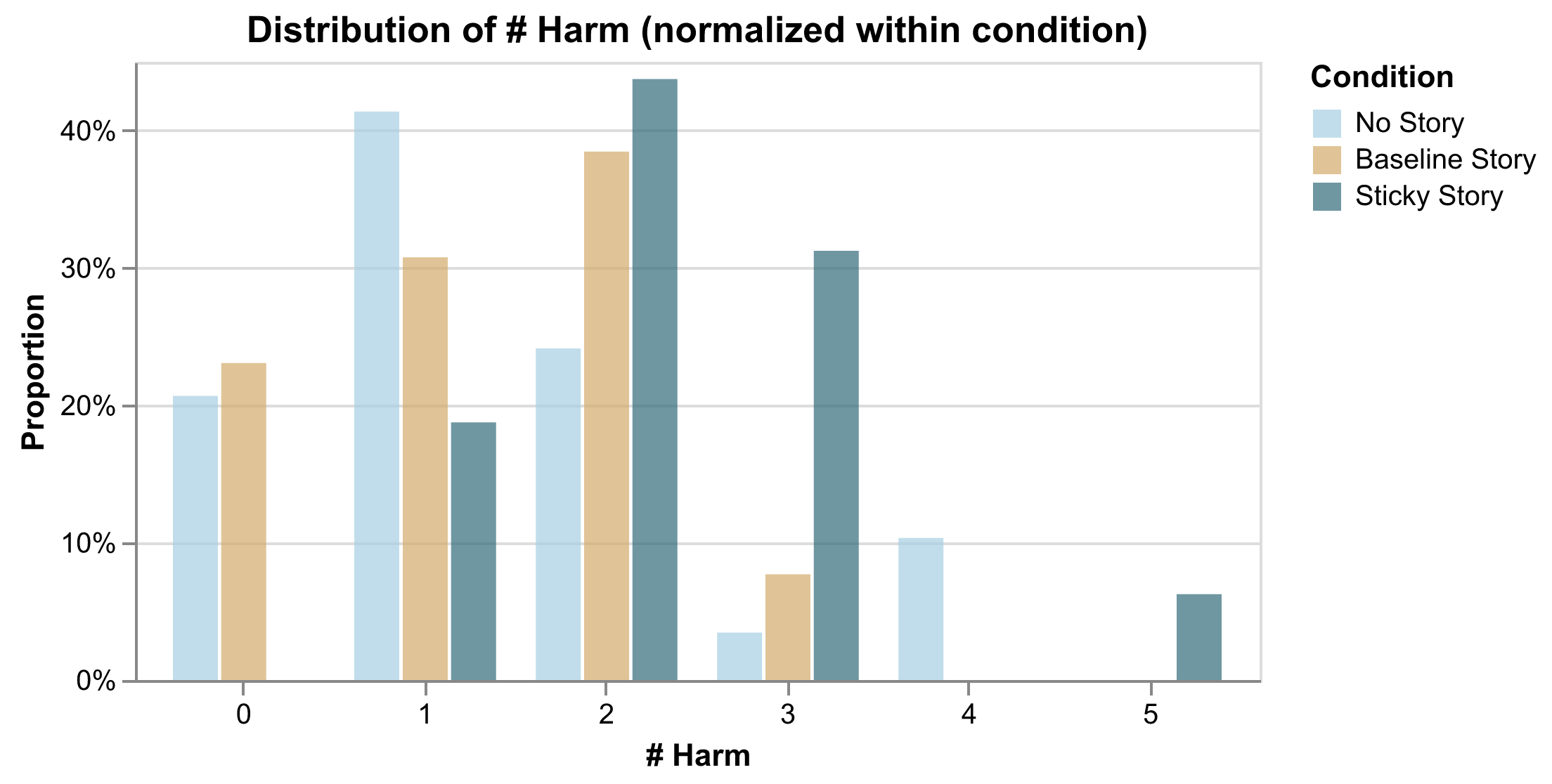}
\caption{The distribution of the number of harms identified before and after the intervention.}
\label{fig:histogram}
\end{figure}
}

\newcommand\figureTemplate[1]{
\begin{figure}[t]
\centering
\includegraphics[width=1\linewidth]{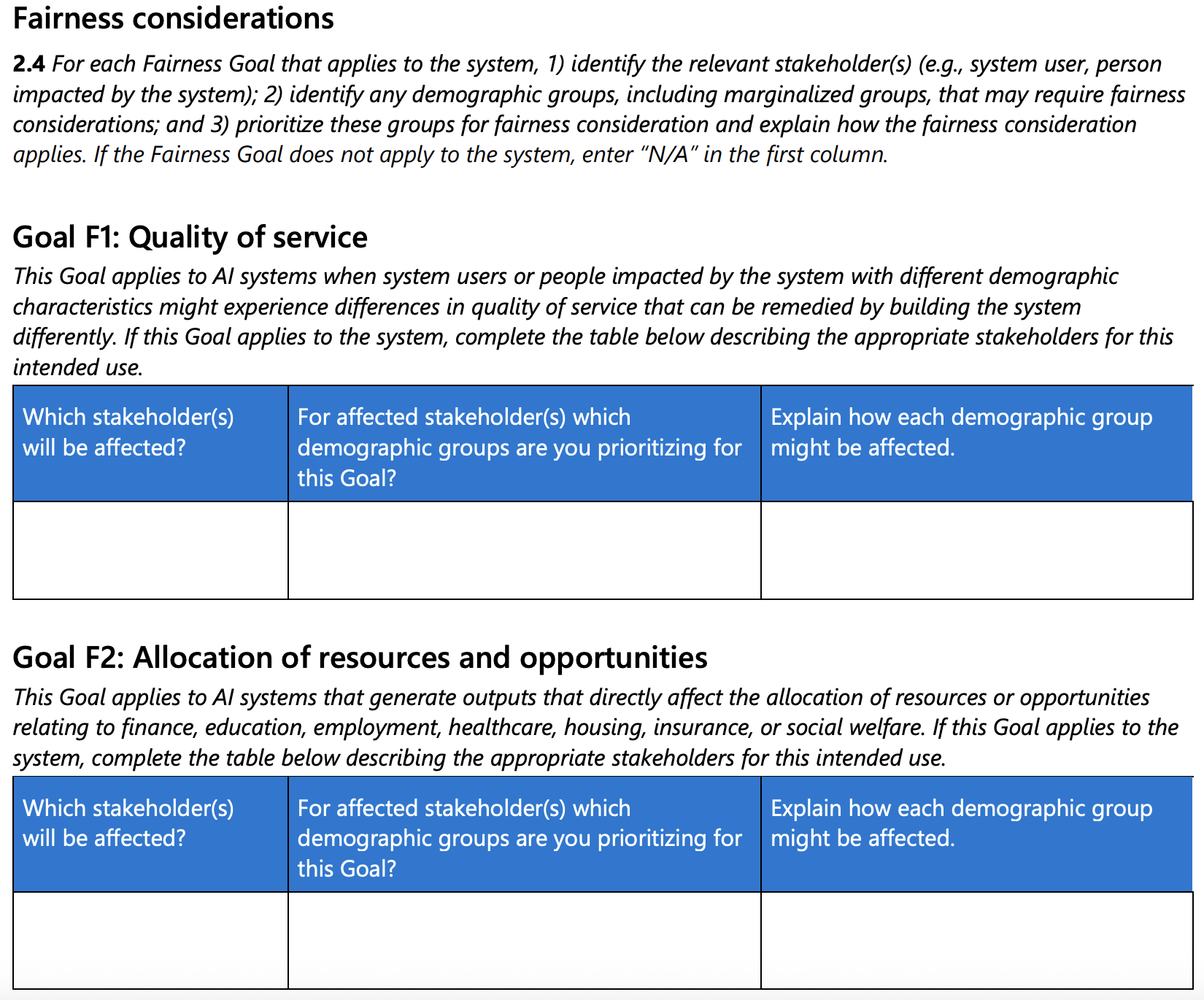}
\caption{#1}
\label{fig:template}
\end{figure}
}

\newcommand\figureStudy[0]{
\begin{figure}[t]
\centering
\includegraphics[width=1\linewidth]{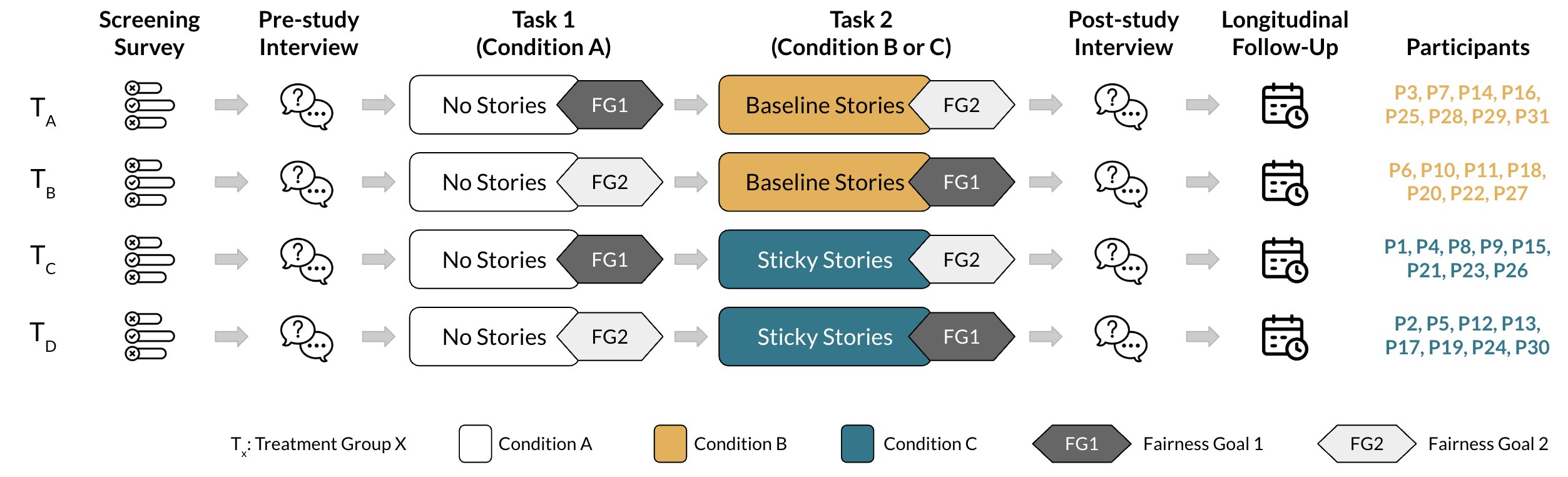}
\caption{The mixed study design for the user study that combines both within-subject and between-subject elements, to evaluate the effectiveness sticky stories compared to no stories and baseline stories}
\label{fig:study}
\end{figure}
}

\newcommand\figureCombined[0]{
\begin{figure}[t]
\centering
\includegraphics[width=0.7\linewidth]{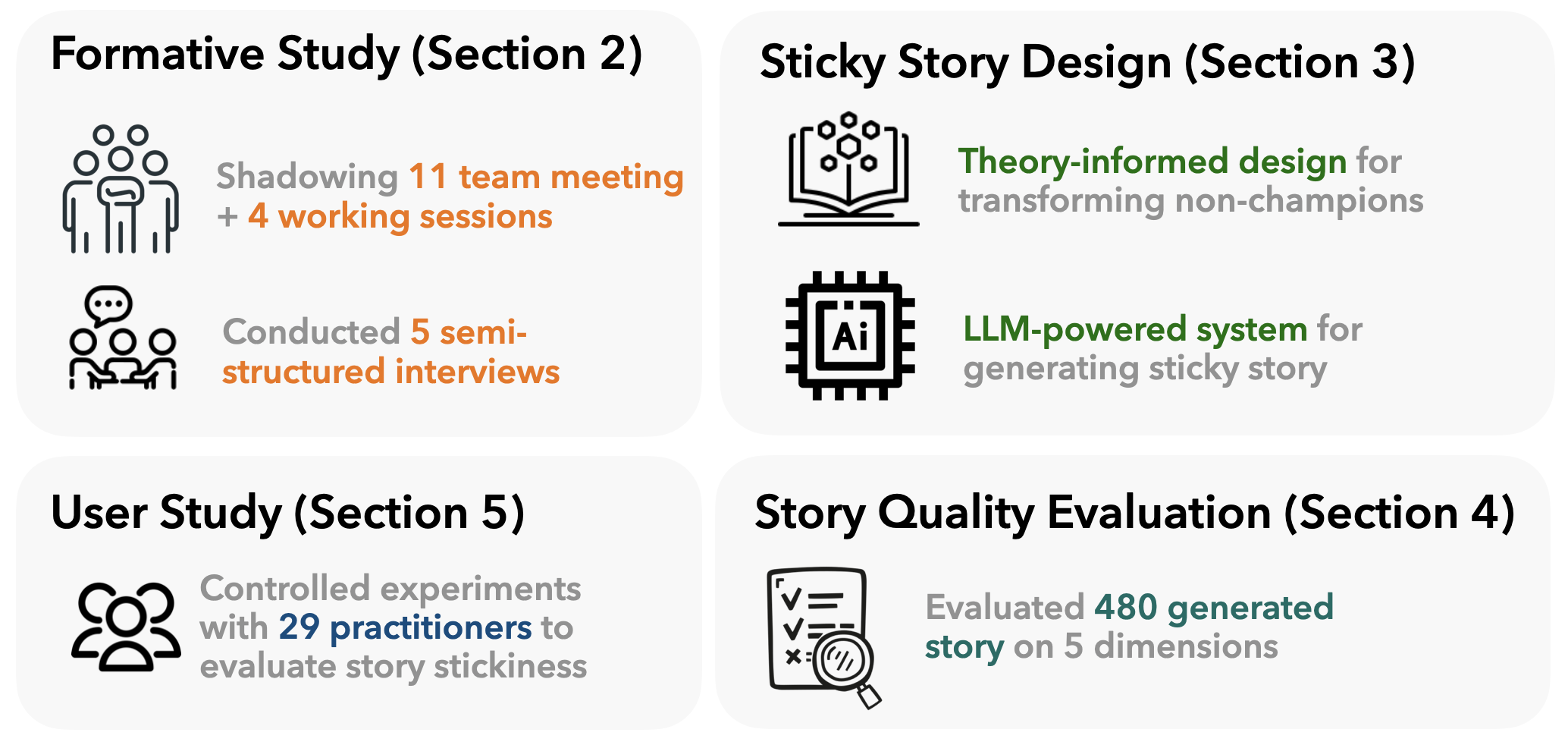}
\caption{Research overview. The work spans four parts: (1) a formative study, (2) a theory‑informed, LLM‑powered design for generating “sticky” stories, (3) a quality evaluation of sticky stories, and (4) a controlled user study.}
\label{fig:combined}
\end{figure}
}

\newcommand\tableOfflineEval[0]{
\begin{table}[]
\caption{Offline evaluation results of the generated stories.}
\label{tab:offlineeval}
\begin{tabular}{@{}lrrrrrrr@{}}
\toprule
 & \multicolumn{5}{c}{\textbf{Quality}} & \multicolumn{2}{c}{\textbf{Cost}} \\ \cmidrule(lr){2-6} \cmidrule(lr){7-8}
 & Severity & Surprising & Concrete & Relevant & Diversity & Token & Time/s \\ \midrule
Pipeline & \textbf{0.992} & \textbf{0.783} & \textbf{1.000} & 0.892 & \textbf{0.156} & 56665 & 50 \\
Baseline & 0.683 & 0.354 & 0.017 & \textbf{0.979} & 0.098 & 1232 & 9 \\
\bottomrule
\end{tabular}
\end{table}
}

\newcommand\tableGQM[0]{
\begin{table}[h]
\centering
\caption{Goal–Question–Metric (GQM) alignment for evaluating practitioner engagement and reasoning.}
\begin{tabular}{p{2.5cm} p{5.5cm} p{4cm} p{2cm}}
\toprule
\textbf{Goal} & \textbf{Question} & \textbf{Metric} & \textbf{Data Source} \\
\midrule
Understand engagement with stories & Do participants identify more potential harm categories and invest more time when exposed to sticky vs baseline stories? & Number of harms flagged, diversity of harms (coded according to harm taxonomy), time spent per task & Tool interaction logs \\
\addlinespace
Understand reasoning processes & How do participants justify harm identification under baseline vs sticky story conditions? & Comparative coding for reasoning patterns & Think-aloud transcripts \\
\addlinespace
Assess reflection and follow-up & Do participants act on insights or discuss them with peers? & Presence/absence of reported actions, types of discussions with peers & Follow-up survey responses \\
\bottomrule
\end{tabular}
\label{tab:gqm}
\end{table}

}

\newcommand\tableAnovaTime[0]{
\begin{table*}[t]
\centering
\small
\begin{tabular}{l|cc|cc|cc}
\toprule
          & \multicolumn{2}{c|}{\textbf{Relative time (tast 2/task 1)}} & \multicolumn{2}{c|}{\textbf{Task 1 time}} & \multicolumn{2}{c}{\textbf{Task 2 time (ANCOVA)}} \\
\midrule
\textbf{Source}       & F     & p     & F     & p     & F     & p \\
\midrule
Story condition (baseline vs sticky)          & 31.37*** & <0.001 & 4.39*    & 0.047 & 28.47*** & <0.001 \\
Task/fairness goal order          & 0.60     & 0.445  & 1.01     & 0.324 & 2.48     & 0.130 \\
RAI awareness score      & 1.07     & 0.313  & 2.13     & 0.158 & 5.11*    & 0.034 \\
Championship score   & 0.03     & 0.855  & 0.20     & 0.656 & 2.81     & 0.108 \\
Prior AI/ML experience     & 0.24     & 0.629  & 0.34     & 0.568 & 0.82     & 0.376 \\
Task 1 time (cov.)  & —        & —      & —        & —     & 18.25*** & 0.0003 \\
\bottomrule
\multicolumn{7}{l}{\small{$^{***}p<0.001,\quad ^{**}p<0.01,\quad ^{*}p<0.05$}} \\
\end{tabular}
\caption{Two-way ANOVA/ANCOVA results for time spent on Task 1 and Task 2. F-values and p-values are reported for all sources. Partial eta squared ($\eta_p^2$) for the main Story condition effect was 0.60 for Relative time, 0.08 for Task 1 time, and 0.33 for Task 2 time. Significant effects are indicated with stars.}
\label{tab:anovatime}
\end{table*}
}

\newcommand\tableAnovaHarm[0]{
\begin{table*}[t]
\centering
\small
\begin{tabular}{l|cc|cc|cc|cc}
\toprule
\multirow{2}{*}{Source} & \multicolumn{2}{c|}{Task 1 tax.} & \multicolumn{2}{c|}{Task 2 tax. (ANCOVA)} & \multicolumn{2}{c|}{Task 1 subtax.} & \multicolumn{2}{c}{Task 2 subtax. (ANCOVA)} \\
\cmidrule(lr){2-3} \cmidrule(lr){4-5} \cmidrule(lr){6-7} \cmidrule(lr){8-9}
 & F & p & F & p & F & p & F & p \\
\midrule
Story condition (baseline vs sticky) & 1.74 & 0.200 & 19.43*** & 0.0002 & 0.73 & 0.403 & 22.39*** & 0.0001 \\
Task/fairness goal order & 0.041 & 0.841 & 0.001 & 0.979 & 0.174 & 0.681 & 0.026 & 0.873 \\
RAI awareness score & 0.057 & 0.813 & 0.663 & 0.424 & 0.000003 & 0.999 & 0.004 & 0.950 \\
Championship score & 0.257 & 0.617 & 3.57 & 0.072 & 0.005 & 0.947 & 0.44 & 0.514 \\
Prior AI/ML experience & 0.119 & 0.734 & 0.000003 & 0.999 & 0.074 & 0.787 & 0.004 & 0.952 \\
Task 1 tax.(cov.) & -- & -- & 18.25*** & 0.0003 & -- & -- & -- & -- \\
Task 1 subtax. (cov.) & -- & -- & -- & -- & -- & -- & 0.018 & 0.896 \\
\bottomrule
\multicolumn{9}{l}{\small{$^{***}p<0.001,\quad ^{**}p<0.01,\quad ^{*}p<0.05$}}
\end{tabular}
\caption{ANOVA and ANCOVA results for taxonomy (tax.) and sub-taxonomy (subtax.) measures. F-values and p-values are reported for all sources. Partial $\eta_p^2$ for the main Condition effect are 0.09 (Task 1 tax.), 0.49 (Task 2 tax.), 0.036 (Task 1 subtax.), and 0.54 (Task 2 subtax.).}
\label{tab:anovaharm}
\end{table*}
}

\newcommand\tableDescriptive[0]{
\begin{table*}[t]
\centering
\small
\begin{tabular}{l l c c c c}
\toprule
\textbf{Condition} & \textbf{Task} & \textbf{Time (min)} & \textbf{Harms} & \textbf{Harm categories} & \textbf{Harm subcategories} \\
\midrule
\multirow{2}{*}{Baseline condition} 
  & Task 1 (no stories)  & 7.5 ± 4.3 & 1.46 ± 1.05 & 1.23 ± 0.73 & 1.38 ± 0.77 \\
  & Task 2 (baseline stories) & 8.3 ± 3.6 & 1.31 ± 0.95 & 0.31 ± 0.48 (↑) & 0.54 ± 0.66 (↑)\\
\midrule
\multirow{2}{*}{Sticky condition} 
  & Task 1 (no stories) & 5.4 ± 2.7 & 1.31 ± 1.30 & 0.81 ± 0.66 & 1.06 ± 0.93 \\
  & Task 2 (sticky stories)& 16.6 ± 7.4 & 2.31 ± 1.01 & 1.38 ± 0.62 (↑) & 1.88 ± 0.62 (↑) \\
\bottomrule
\end{tabular}
\caption{Descriptive statistics for time on tasks, number of harms identified, and diversity of harms (categories and subcategories) across conditions. Values are reported as mean ± standard deviation (SD). Task 2 “Harm categories” and “Harm subcategories” indicate additional counts relative to Task 1 (↑)}
\label{tab:descriptive}
\end{table*}
}

\newcommand\figuretimeChangeChart[0]{
\begin{figure}[t]
\centering
\includegraphics[width=\linewidth]{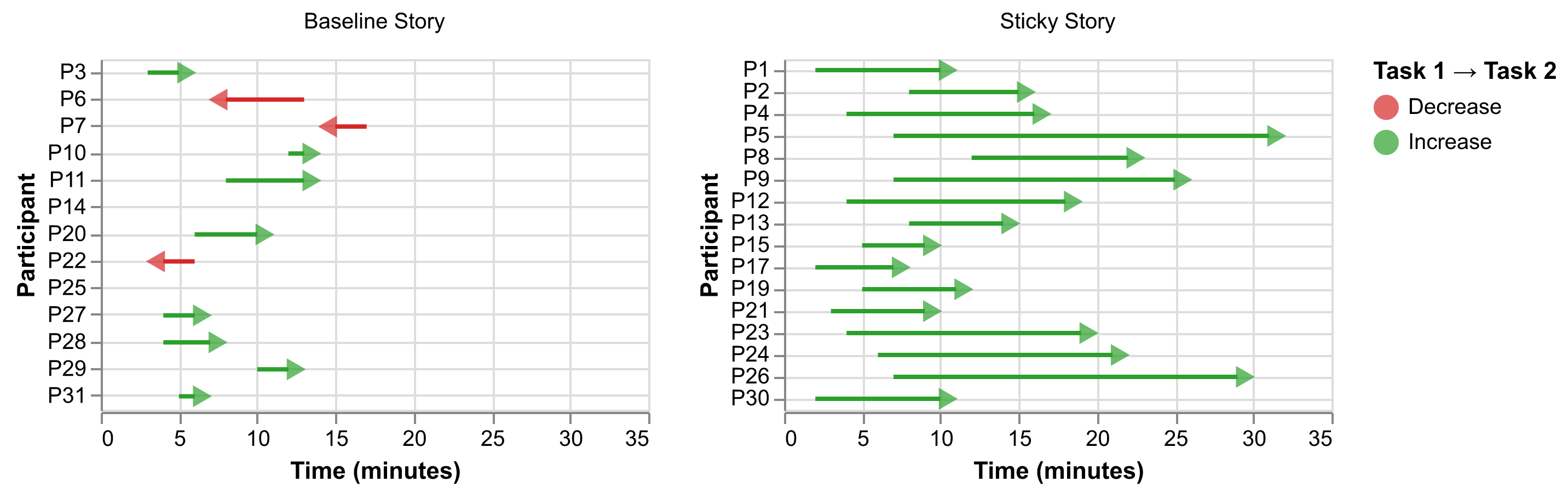}
\caption{We measured the time participants spent on the harm identification task 1 (no stories) and task 2 (baseline or sticky stories). We found that participants who read sticky stories consistently spent more time on  the harm identification task, with much higher increases.
}
\label{fig:timeChangeChart}
\end{figure}
}

\newcommand\tableReflection[6]{
\begin{table}[h!]
\centering
\footnotesize
\caption{Signs of Critical Reflection in Non-Champions}
\label{tab:criticalreflection}
\begin{tabular}{@{}p{1.5cm} p{4cm} p{4cm} p{4.8cm}@{}}
\toprule
\textbf{Sign} & 
\textbf{Description} & 
\textbf{Indicators} & 
\textbf{Example} \\ 
\midrule
#1 & Practitioners critically examine or dispute underlying beliefs, premises, or taken-for-granted assumptions that might otherwise go unquestioned. & Identifies own beliefs that may not hold; contrasts story assumptions with their own understanding, experience, or evidence; expresses skepticism toward “default” ways of thinking or doing things. & \PTwentyThree: \textit{"I think human evaluators are really important -- it's something that I am kind of understanding now. We haven't done this. But looking at the stories, I think, it's a really important for us" }
\\
#2 & Practitioners consider scenarios from multiple angles or stakeholders, compare alternatives, or expand the scope beyond their immediate perspective. & Weighing different interpretations; comparing different parts of a story; acknowledging multiple voices. & \PFifteen: \textit{"As a data scientist, I can retrain the model, and I can test it. But what can a government or the person in FDA do? They don't know this."} \\
#3 & Practitioners connect the story to broader organizational, social, or technical systems, often moving beyond the immediate prompt. & Linking story implications to other domains, contexts, or systemic risks; recalling real-world events. & \PFour: \textit{"[...] like scholarship distribution and funding distribution specifically, whatever was coming to any charity. How are they distribute it amongst like schools or old age homes and other organizations where the funding has to go to."} \\
#4 & Practitioners show surprise, novelty, or realization about aspects of the scenario, signaling a shift in understanding. & Expressions of being surprised, not having thought of it before, finding something new, unexpected, or revealing. & \PNineteen: \textit{"oh, that's mind blowing [...] I didn't know this can happen."} \\
#5 & Practitioners demonstrate a back-and-forth reasoning process, revising or expanding their views as they talk. & Evidence of reconsideration, self-correction, extended reflection, or stepwise elaboration. & \PTwentySix: \textit{"Now that I'm looking at the word like demeaning. I feel like I can probably think of a couple more examples."} \\
#6 & Practitioners translate reflection into concrete plans for action or acknowledge the need to modify future behavior, processes, or designs. & Commitment to follow-up action; expressing intent to discuss with others; specific takeaways for their own projects. & 
\PThirty: \textit{"I think, filtration of the training data is pretty simple. It could be done with a simple  Regex based filtration, for at least getting rid of such harmful potential comments. And post filtration for the same should be simple, too. Yeah. So I think that's pretty painless to deploy. I would be motivated to do that."}\\
\bottomrule
\end{tabular}
\end{table}
}

\newcommand\tableVariable[2]{
\begin{table}[ht]
\centering
\caption{Study variables with type, operationalization, and data source. Icons indicate data source: \faPen{} = self-reported, \faCalculator{} = derived, \faTags{} = coded.}
\label{tab:variables}
\begin{tabular}{p{2.5cm} p{3cm} p{7cm} p{1cm}}
\toprule
\textbf{Type} & \textbf{Variable} & \textbf{Operationalization} & \textbf{Source} \\
\midrule
Independent & Story condition & No story (A) vs. baseline story (B) vs. sticky story (C) & – \\
\midrule
Dependent (Quant.) & Time spent on harm identification & Duration of engagement, extracted from tool logs & \faCalculator \\
Dependent (Quant.) & Number of harms & Number of identified harms, extracted and counted from tool logs & \faCalculator \\
Dependent (Quant.) & Distinct harm categories & Number of unique categories/subcategories identified, coded with RAI harm taxonomy #1 & \faTags \\
Dependent (Qual. \& Quant.) & Engagement behaviors & Frequency of coded signs of critical reflection (#2), from think-aloud transcripts & \faTags \\
Dependent (Qual.) & Self-reported planned actions & Reported and intended actions from follow-up survey responses & \faPen \\
Dependent (Qual.) & Transformation trajectories & Shifts in perspective, identified from transcripts and video recordings & \faTags \\
\midrule
Control & RAI awareness score & 0–2 scale, based on in-session remarks:\\
 & & 0 = unaware, 1 = partially aware, 2 = aware & \faTags \\
Control & Championship score & 0–5 scale of advocacy for RAI principles in work \\
 & & 0 = unaware, 1 = actively opposed, 2 = acknowledges importance but takes no action, 3 = follows processes reluctantly, 4 = willing to contribute but not advocating, 5 = actively promotes RAI practices among peers & \faTags \\
Control & Prior AI/ML experience & Self-reported in pre-screening survey: \\  & & 1-2 years, 3-5 years, 6-10 years, more than 10 years & \faPen \\
Control & Task/fairness goal order & Randomly assigned & - \\
\bottomrule
\end{tabular}
\end{table}
}

\begin{document}
\looseness=-1
\title{``I Don't Think RAI Applies to My Model'' -- Engaging Non-champions with Sticky Stories for Responsible AI Work}
\author{Nadia Nahar}
\email{nadian@andrew.cmu.edu}
\affiliation{%
   \institution{Carnegie Mellon University}
   \country{USA}}
   \author{Chenyang Yang}
\affiliation{%
   \institution{Carnegie Mellon University}
   \country{USA}}
   \author{Yanxin Chen}
\affiliation{%
   \institution{Carnegie Mellon University}
   \country{USA}}
   \author{Wesley Hanwen Deng}
\affiliation{%
   \institution{Carnegie Mellon University}
   \country{USA}}
   \author{Ken Holstein}
\affiliation{%
   \institution{Carnegie Mellon University}
   \country{USA}}
   \author{Motahhare Eslami}
\affiliation{%
   \institution{Carnegie Mellon University}
   \country{USA}}
   \author{Christian Kästner}
\affiliation{%
   \institution{Carnegie Mellon University}
   \country{USA}}
\renewcommand{\shortauthors}{N. Nahar et al.}

\begin{abstract}

\figureOverview
Responsible AI (RAI) tools—checklists, templates, and governance processes—often engage RAI champions, individuals intrinsically motivated to advocate ethical practices, but fail to reach non-champions, who frequently dismiss them as bureaucratic tasks. To explore this gap, we shadowed meetings and interviewed data scientists at an organization, finding that practitioners perceived RAI as irrelevant to their work. Building on these insights and theoretical foundations, we derived design principles for engaging non-champions, and introduced sticky stories—narratives of unexpected ML harms designed to be concrete, severe, surprising, diverse, and relevant, unlike widely circulated media to which practitioners are desensitized. Using a compound AI system, we generated and evaluated sticky stories through human and LLM assessments at scale, confirming they embodied the intended qualities. In a study with 29 practitioners, we found that, compared to regular stories, sticky stories significantly increased time spent on harm identification, broadened the range of harms recognized, and fostered deeper reflection.
\end{abstract}
                \maketitle

        \section{Introduction}\label{h.odjbzzdx44bn}
\emph{``David was the only employee who used the building's accessibility elevator. When the company deployed predictive maintenance AI, the algorithm learned his elevator had extremely low usage and began delaying its repairs to prioritize busy main elevators.}

\emph{David's elevator grew unreliable, but his complaints were ignored---the system marked it "low priority." During the annual shareholder meeting, it broke down completely, trapping him for hours while VIPs toured the building.}

\emph{David shared his experience on social media, and the story quickly went viral. Local news stations picked up the story, framing it as "AI bias leaves disabled worker trapped." The coverage triggered an ADA investigation revealing the AI had systematically neglected accessibility infrastructure based on usage metrics. Worse, the same system was deployed across dozens of buildings, all with severely under-maintained accessibility elevators now requiring emergency repairs. The company faced millions in fines, settlements, and unexpected maintenance costs.''}

We open with this LLM-generated vignette to make a point central to this paper: Surprising stories relevant to our own work cut through attention fatigue. As we will show, practitioners in ``non‑critical'' domains are often desensitized by a steady stream of algorithmic bias headlines (e.g., biased hiring algorithms \gencite{bGWG}{[15]}{}, bias in predictive policing and recidivism assessment \gencite{EDUM,E7oK}{[62, 104]}{}, autonomous vehicle crashes \gencite{5Jye}{[13]}{}) that seem bad but not relevant to them; as a result, fairness discussions remain abstract and easy to deprioritize. Despite a surge of checklists \gencite{kBDV}{[45]}{}, templates \gencite{AwKU,lem8}{[27, 56]}{}, games \gencite{ntG3,9xTg,7Jwy,qAIi,6v2p}{[3, 39, 49, 87, 89]}{}, and governance processes \gencite{99HI}{[70]}{} for Responsible AI (RAI), prior research \gencite{6Tr2,kBDV,V4MU,Ufiz,99HI}{[32, 45, 65, 70, 71]}{} has primarily examined their effects on \emph{RAI champions}---individuals intrinsically motivated to advocate for ethical practices within their organizations (sometimes in formalized roles \gencite{vnKH}{[88]}{})). Much less is known about the majority of practitioners---\emph{non-champions}, those without prior motivation or formal RAI roles---and whether they would meaningfully engage with these resources or integrate them into their work. In this paper, we focus on \emph{engaging non-champions to deliberate about RAI}. 

In a formative study with a partner technology organization---where we shadowed meetings and interviewed governance members and data scientists---we observed a sharp contrast: Governance champions actively advanced RAI and designed governance structures, whereas many data scientists remained dismissive and disengaged regarding RAI concerns. Even when required to participate in activities, non-champions checked the boxes with minimal engagement. The key bottleneck was not the availability of guidance, processes, or tools, but the challenge of motivating these practitioners to meaningfully engage. This gap is critical, as RAI initiatives and tools cannot achieve impact if practitioners do not engage with them in a meaningful way. This observation motivated our research question: \textbf{How can we design interventions that foster deep engagement among non-champion practitioners in RAI processes?} By \emph{deep engagement}, we mean more than just ticking boxes or going through the motions---it involves practitioners \emph{critically reflecting} on potential harms, considering trade-offs thoughtfully, and retrospecting on their past experiences.

In this paper, we introduce an intervention designed to specifically engage \emph{non-champion} practitioners by presenting them narrative-based scenarios, like the elevator-maintenance one above, that illustrate possible \emph{unexpected} and \emph{severe} real-world harms arising from \emph{their own} ML systems. Drawing on theories from psychology and business communication  \gencite{JO4x,bdaT}{[31, 51]}{}, we developed an LLM-based system that generates scenarios to embody five qualities that are known to drive engagement and memorability: Scenarios should be \emph{concrete}, \emph{severe}, \emph{surprising}, \emph{diverse}, and \emph{relevant.} Inspired by the framework in \emph{Made to Stick \gencite{JO4x}{[31]}{}}, we refer to these narrative scenarios as \textbf{sticky stories}. Unlike conventional documentation \gencite{lem8,jLLo}{[55, 56]}{} or checklists \gencite{kBDV}{[45]}{} or generic vignettes \gencite{vbPE}{[11]}{}, which often feel abstract or disconnected from the realities of product development, our sticky stories are designed to evoke curiosity and spark critical reflection. 

To evaluate the effectiveness of our \emph{sticky stories}, we conducted two complementary evaluations. First, we assessed the ``\emph{stickiness}'' of the generated stories themselves, that is, whether they embodied the five key qualities. The results showed that \emph{sticky stories} significantly outperformed stories generated by zero-shot prompts used in past work across most qualities (e.g., severity: 99\% vs 68\%, and surprisingness 78\% vs 35\%) (cf. table~\ref{tab:offlineeval}). Second, we conducted a user study with 29 practitioners, mostly non champions, to measure the practical impact of these stories on engagement. Practitioners exposed to sticky stories spent significantly more time on harm identification (10 times more), identified a larger number of harm categories and subcategories (five times as many new categories and 3.5 times as many subcategories), and engaged in deeper critical reflection compared to those seeing baseline stories. Practitioners also exhibit distinct trajectories in shifting their attitudes from initial indifference or resistance toward a more engaged stance on RAI.

\paragraph{Contributions.}
 This work makes the following contributions (see the study overview in Fig.~\ref{fig:combined}). 

\begin{compactitem}
	\item Our formative study characterizes the engagement gap among non-champion practitioners with Responsible AI efforts, and shows that existing governance tools and templates are insufficient to motivate meaningful participation. 

	\item The design of \emph{sticky stories} based on key qualities that capture practitioners' attention, provoke reflection, and foster engagement.

	\item A scalable compound AI system to generate stories that meet these qualities and integrate them into an interactive tool for practitioner use.

	\item An empirical evaluation demonstrating that sticky stories increase engagement time, the number and diversity of harms identified, and promote critical reflection.

	\item Practitioner-specific engagement trajectories, highlighting how different champions and non-champions in different profiles  respond differently.

\end{compactitem}

\section{Related Work}\label{h.mrie1xvfofs5}
Early work seeking to understand industry RAI practices suggested that RAI efforts were often driven by ``individual advocates'' who are self-motivated to pursue RAI work \gencite{kBDV,6Tr2}{[45, 71]}{}. In recent days, many practitioners are now formally tasked by their organizations with considering RAI issues \gencite{Gyex,9tBg,gFFd}{[47, 78, 95]}{}. In this paper, we use ``\emph{RAI champion},''  a term used by organizations such as Microsoft as a role title \gencite{FFVF}{[75]}{}, to refer to both self-motivated advocates and formally designated and trained RAI roles. While there is an abundance of prior work focusing on challenges RAI champions face and how to support them, it remains unclear whether such approaches transfer to the broader group of practitioners that we refer to as \emph{non-champions}, namely those who are not already motivated to lead RAI work but nonetheless encounter RAI concerns in their everyday roles.

\subsection{State of Responsible AI in Industry}\label{h.7w70affw492k}
\paragraph{Prior research has extensively examined industry RAI practices and challenges.}
 Within the CHI and broader HCI communities, there has been a strong push to better understand industry RAI practices, challenges, and needs \gencite{V4MU,wb7X,6Tr2,Ufiz}{[32, 65, 71, 86]}{}. For instance, through interviews and surveys, Holstein et al. \gencite{V4MU}{[32]}{} identified challenges in fairness-aware data collection and introduced proactive auditing processes. Focusing on UX professionals, Liao et al. \gencite{1IA0}{[43]}{} and Wang et al. \gencite{3qzG}{[96]}{} emphasized the need for improved tools and prototyping methods to facilitate communication and collaboration with technical teams when addressing RAI concerns. 

\paragraph{Prior research highlights organizational challenges and risks that can limit the meaningful implementation of responsible AI in industry.}
 A large body of HCI and RAI research has shown that individuals frequently encounter pushback from leadership when advocating for more responsible technologies \gencite{ShxA,hLof,wb7X,wItU,IuY2,LX8x}{[1, 5, 36, 86, 100, 101]}{}. In addition, despite the many RAI principles, guidelines, and frameworks published by technology companies, organizational studies of industry RAI practices have consistently highlighted how the profit-driven and fast-paced nature of industry work often demotivates practitioners from engaging in meaningful RAI efforts \gencite{9tBg,Ufiz,jrdw,hnb8}{[47, 65, 82, 99]}{}. As a result, multiple studies warn that RAI processes risk becoming bureaucratic ``check-the-box'' exercises rather than reflective, substantive practices \gencite{kBDV,9zRX,Wppy,p90f,K0Qs}{[9, 40, 45, 93, 94]}{}. For instance, RAI documentation is often reduced to a compliance task \gencite{kBDV,adh7,249M,iS9g}{[12, 19, 45, 102]}{}, while fairness and explainability evaluations can become performative practices, sometimes criticized as ethics washing \gencite{KUIg,eNFn,ccoc,C9JW}{[2, 17, 46, 77]}{}.

\subsection{Supports for RAI Harm Identification}\label{h.chf9ecn0jmb3}
\paragraph{An abundance of structured templates and frameworks exist to support practitioners in Responsible AI impact assessment.}
 To support practitioners in identifying potential harms and ethical risks, a wide range of RAI assessment approaches have been introduced, often in the form of structured checklists or impact assessments. For example, Microsoft's \emph{RAI Impact Assessment Template} provides guidelines for conducting impact reviews prior to deploying AI products  \gencite{jLLo}{[55]}{};  Bogucka et al. \gencite{yqHL}{[7]}{} co-designed and evaluated an AI impact assessment template with practitioners and compliance experts grounded it in regulatory requirements;  Deng et al. \gencite{spJu}{[16]}{} developed a \emph{Societal Impact Assessment} template focused on design considerations for effective adoption and adaptation; and Rismani et al. \gencite{FdBE}{[79]}{} argue for the use of established hazard engineering techniques to structure the analysis

More broadly, several tools aim to promote broader reflection on the consequences of technology. Nathan et al. \gencite{kkYb}{[60]}{} developed a tool to help practitioners envision long-term effects of interactive systems. Elsayed-Ali et al. \gencite{2Fiv}{[22]}{} introduced \emph{Responsible \& Inclusive Cards}, an online card-based tool designed to encourage critical reflection on project impacts. Ehsan et al. \gencite{xACN}{[21]}{} introduced \emph{Seamful XAI} to allow stakeholders identify mistakes and enhance AI explainability. Documentation frameworks such as \emph{Datasheets for Datasets} [31], originally intended to improve transparency in data collection, have also been shown to help surface ethical concerns [11].

This line of work emphasizes structured processes and tooling as a means to support developers in  anticipating harms and fostering reflection. 

\paragraph{Recent research has begun leveraging large language models (LLMs) to help AI practitioners reflect on potential risks and harms in their systems.}
 Building on prior HCI work that demonstrated the potential of large language models (LLMs) to support brainstorming and reflexivity, researchers have begun exploring how to incorporate LLMs into RAI tools to support reflection around RAI concerns \gencite{vbPE,Func,73Qw}{[11, 64, 97]}{}. Approaches either (a) use LLMs to generate examples of possible harms for a system, following structured reasoning internally to create diverse harms, such as the vignettes generated by  AHA! \emph{\gencite{vbPE}{[11]}{}} and our own work in this paper, or (b) identify and present real-world reports about related systems from news stories as in Farsight \gencite{73Qw}{[97]}{} and BLIP \gencite{Func}{[64]}{}. Both strategies aim to guide analysis with realistic examples and broaden the range of consequences considered.

We find these tools promising and build on similar ideas. While their design may not explicitly target RAI champions, we suspect that they are more effective for developers already motivated for RAI work. Since motivated RAI practitioners are more likely to volunteer for evaluations of RAI tools (self-selection bias) and the participants' prior motivation was not controlled for in prior studies, we are curious about how effective such tools are for a broad range of practitioners.

\subsection{Engaging Non-Champions}\label{h.rfsjq06do79w}
\paragraph{Research suggests that existing fairness and interpretability tools often fail to foster genuine engagement, instead encouraging superficial compliance and limiting meaningful understanding.}
 As some organizations make RAI steps mandatory, either through explicit RAI audit gates \gencite{HQEK,Clqd,pZ76}{[73, 74, 76]}{} or by attaching RAI considerations to existing required privacy assessment steps \gencite{GsMU}{[18]}{}, it remains to be seen whether non-champions engage in depth or just do the minimum amount of work to complete the necessary steps (``check-the-box compliance'')

Research has found evidence of such a check-the-box culture: For instance, a participant in Balayn et al.'s study noted, ``\emph{Fairness for many companies is just a small checkbox, and sometimes people put their mark without any question…}'' \gencite{C9JW}{[2]}{}. Similarly, Kaur et al. \gencite{JRGk}{[37]}{} found that interpretability tools, while designed to improve understanding of machine learning models, can sometimes impair it, with strategies aimed at promoting deliberation and engagement frequently failing to overcome this, and Omar et al. \gencite{TnBR}{[61]}{} found that structured policy guidance was not effective at engaging with user needs for explanation designs.

\paragraph{Recent research has begun exploring strategies to involve non-champions in responsible AI work, though significant gaps remain.}

Common strategies to attempt to engage practitioners for RAI work involve nudging \gencite{8K4v}{[6]}{}, gamification \gencite{ntG3,9xTg,7Jwy,qAIi}{[3, 39, 87, 89]}{}, and reframing fairness work in familiar quantitative terms \gencite{GsMU}{[18]}{}. For example, Bhat et al. introduced a JupyterLab extension to nudge data scientists to complete and update model card documentation, particularly the ethics-related sections  \gencite{8K4v}{[6]}{}. Ballard et al. proposed \emph{Judgment Call}, which helps product teams surface ethical concerns using value-sensitive design and design fiction \gencite{ntG3}{[3]}{}, and Kim et al. developed \emph{The Desk: Dilemmas in AI Ethics}, a digital game-based approach to enhance learning about AI ethics \gencite{9xTg}{[39]}{}. 

While these strategies can capture short-term attention, it is not clear that they are sustainable to keep non-champions engaged. Nudges may increase initial actions, but cause longer-term behavior changes---and may even reduce follow-through in some cases \gencite{jNpH,eJsl,rz6N}{[34, 67, 105]}{}. Similarly, gamification can spike early engagement, yet motivation drops as the novelty wears off, and in some contexts may impair deeper learning or distract from authentic engagement \gencite{oPti,4XW6}{[30, 80]}{}. For example, Widder et al. \gencite{hnb8}{[99]}{} argued when evaluating one of these games, that hypothetical contexts created in the game are unlikely to be a viable mechanism for real world change.

In summary, prior research has explored barriers to RAI work and provided many processes and tools to support and engage practitioners in RAI work. However, as we have experienced and will describe next, this support is not equally effective for all practitioners and may fall short for non-champions who are not motivated to go beyond minimally required steps, if any. To the best of our knowledge, prior research focuses mostly (deliberately or not) on RAI champions and has not explored the differences between champions and non-champions. In this paper, we aim to fill this gap, by focusing specifically on how to engage non-champions for RAI activities.

\figureCombined

\section{Phase I: Formative Study }\label{h.wd4r2a1wc5ky}
We conducted a formative study in close collaboration with a partner technology organization seeking the adoption of Responsible AI (RAI) processes organization-wide. Our initial goal was to explore how best practices could be introduced to strengthen existing RAI governance structures---best practices, such as templates, checklists, and audit processes. However, as the study unfolded, we found that the more pressing challenge was not the refinement of these tools, but the underlying disconnect and resistance among practitioners in engaging with them.

\subsection{Research Method}\label{h.inctudrhrin9}
\paragraph{Data Collection.}
 Our data collection followed a qualitative, ethnographically informed approach, aiming to capture both organizational-level discussions and on-the-ground technical practices. We combined two complementary strategies: (1) shadowing internal governance and project team meetings, and (2) follow-up conversations and semi-structured interviews with team members. Shadowing involved three governance team meetings (April--May 2023), eight project team meetings (July--October 2023), and four one-on-one working sessions (August--November 2023) between a governance team member and a data scientist. To supplement these observations, we conducted two informal one-on-one conversations with developers and five semi-structured interviews: two with governance team champions and three with data scientists from different project teams. Our documentation relied on contemporaneous note-taking, which foregrounded participants' language, emergent themes, and key tensions, while necessarily prioritizing synthesis over verbatim capture. Across these sessions, we accumulated roughly 22 hours of observation and produced over 50 pages of detailed field notes documenting interactions and discussions. 

\paragraph{Data Analysis.}
 Since our goal was to produce rich, context-sensitive insights that directly informs the design of interventions, rather than coding every interaction in a traditional thematic analysis, we adopted an interpretive approach \gencite{TWMl,K10U}{[91, 92]}{}: We examined patterns, recurring tensions, and illustrative examples across meetings, interviews, and conversations to understand how RAI practices were enacted and experienced. 

\paragraph{Limitations.}
 Observations were limited to meetings and working sessions, which constrained our view of day-to-day development practices and informal conversations where RAI may have been discussed differently. Second, because we were not permitted to record, our analysis relied on contemporaneous notes that emphasize synthesis rather than verbatim detail, which may have filtered nuance. Third, as with any qualitative analysis, researchers' interpretation shaped how themes were identified and articulated; despite best efforts at collaborative review and discussion among multiple researchers, we cannot entirely exclude the possibility of bias. Finally, our findings are based on a single organizational setting, and while we anecdotally heard similar themes from colleagues in other organizations, the readers should take care when generalizing the findings beyond our setting.

\subsection{Key Finding: Despite Substantial Governance Efforts, RAI Had Minimal Impact, with Practitioners Largely Ignoring or Dismissing RAI Concerns}\label{h.f2hq3mhfaqr3}
Even with well-structured governance processes in place, non-champion practitioners largely failed to engage meaningfully with RAI practices. Our observations, working sessions, and interviews revealed three interconnected patterns that explain this lack of engagement:

\emph{Consistent with prior research findings \gencite{6Tr2,VXnz,Aj36}{[58, 59, 71]}{}, practitioners in the organization perceived RAI as abstract and peripheral.} While the governance team invested significant effort into artifacts (e.g., templates, workflows, and training modules) with the intention of making RAI practices actionable (such as providing instructions and a step-by-step breakdown of tasks), practitioners often were not sure why these practices mattered for their work. For example, a data scientist completing a system design template skipped sections on fairness and validation, focusing only on tracing data sources that supported her immediate task. In project meetings where the governance team was invited, RAI topics were virtually absent, mentioned only once in passing, and consistently treated as low-risk compared to client deadlines. RAI assessment templates were largely ignored. Interviews reinforced these patterns across teams: One developer explicitly mentioned that she would only use the RAI assessment templates if mandated. Another team completed the templates retrospectively solely to demonstrate compliance to clients. Across these instances, RAI was seen as extra paperwork rather than a tool for identifying or managing ethical risks.

\emph{Practitioners outside the dedicated} governance \emph{team were largely dismissive of RAI concerns, even when directly prompted.} While practitioners rarely engaged with RAI practices on their own, explicit inquiries also led to dismissive responses. For example, after several project meetings, when we asked a data scientist about potential RAI concerns for their project, they immediately responded, ``\emph{there are no responsible AI concerns for this project},'' without further reflection. We conjecture that this dismissiveness is partly driven by narrow media narratives, which frame fairness primarily around gender and race. Consequently, this may have reinforced a mindset that RAI concerns are limited to these areas, so when a project did not directly involve gender or race, practitioners overlooked and dismissed other important risks, such as privacy violations, safety issues, algorithmic bias in other demographic groups, entirely without deliberation.

\emph{Governance team efforts, though substantial, fell short of effectively motivating practitioner participation in RAI.} Governance team members were highly supportive, providing active coaching, clearly explaining RAI goals and processes, and guiding practitioners through training materials grounded in academic and industry best practices. They proactively monitored progress, clarified doubts, and encouraged reflection on potential RAI risks. The organization even enlisted external support---including our team---to further enhance these interventions. Despite this intensive effort, however, data scientists remained disengaged, completing only minimal tasks to ``check-the-box'' when directly asked by their managers and resisting prescribed processes. This underscores how carefully designed and actively supported governance mechanisms still require practitioner buy-in.

These observations led us to rethink our approach, considering how to foster genuine engagement with governance practices, rather than simply designing more mechanisms or forced compliance.

\section{Phase II: Sticky Story Intervention Design}\label{h.kmjsct18mtz}
Our goal is to motivate ``non-champions'', that is stakeholders who may be indifferent or skeptical, to meaningfully engage in RAI efforts by making the consequences of neglecting its principles feel real and relevant. In this section, we discuss established theories about how practitioners' attitudes and engagement with RAI can be shifted or transformed, and how these theories inform the design decisions behind our intervention.

\subsection{Theoretical Background and Design Principles}\label{h.hnpapdbw60vp}
\paragraph{Transforming Non-Champions.}
 Our initial aim was to motivate non-champions to care about RAI, essentially \emph{transforming} them into champions. To guide this effort, we turned to \emph{Transformative Learning Theory} \gencite{pdsS}{[53]}{}, which explains how people change deeply held beliefs through \emph{critical reflection}. Central to \emph{Transformative Learning Theory} is the concept of \emph{disorienting dilemma}---an event, challenge, or scenario that disrupts existing assumptions and compels self-examination. This reflective process allows individuals to question their prior beliefs and gradually integrate new perspectives into their mental models. In practice, such transformation is difficult and often slow, requiring interventions that actively provoke reflection rather than passive exposure \gencite{8Ote,j69C}{[54, 98]}{}. \emph{Cognitive Dissonance Theory \gencite{qbBY}{[24]}{}} complements \emph{Transformative Learning Theory} by describing the psychological discomfort (\emph{dissonance}) that arises when new information conflicts with existing beliefs or behaviors. This discomfort can motivate individuals to reflect and reconcile inconsistencies, which may lead to shifts in perspective or behavior.

Together, \emph{Transformative Learning Theory} and \emph{Cognitive Dissonance Theory} suggest that effective interventions must both present compelling challenges to existing assumptions and create opportunities for structured reflection. Prior research in domains such as ethics education, clinical training, and diversity initiatives demonstrates that reflective prompts, scenario-based exercises, and facilitated discussions can successfully trigger these processes  \gencite{aIsR}{[48]}{}. For example, in health professional education, curricula designed around transformative learning principles---including disorienting dilemmas, reflective prompts, and facilitated discussions---have been shown to improve practitioners' ethical reasoning, critical reflection, and readiness to navigate complex real-world challenges \gencite{PRoT}{[81]}{}. Similarly, in higher education, faculty engaged in action research projects grounded in transformative learning reported meaningful changes in their teaching practices and approaches to student learning \gencite{ox3r}{[28]}{}. However, lasting mindset change requires more than a single exposure---it depends on repeated reinforcement, supportive contexts, and engaging formats \gencite{8Ote,RJM7}{[66, 98]}{}. The central question is how to design interventions that provoke these mechanisms to trigger such a transformation among non-champions in RAI practices.

Our aim is to trigger such a transformation, not just "tricking" professionals to engage in the short term (e.g., with nudging or gamification). Therefore, we sought to create moments of \emph{disorienting dilemmas} to cause \emph{cognitive dissonance}---points of discomfort strong enough to disrupt taken-for-granted assumptions. However, not all disruptions are equally effective. Our formative findings revealed that practitioners had become desensitized to widely circulated media narratives of bias framed around gender and race, often dismissing these as irrelevant to their own projects. To overcome this resistance, we instead surfaced dilemmas that were likely both unfamiliar to the practitioners and directly consequential within their own work contexts.

\paragraph{Sticky Stories as Intervention.}
 One way to capture practitioners' attention is by showing them stories of harm caused by their own ML systems. Prior work has demonstrated that vignettes and fictional scenarios can be effective for RAI champions \gencite{vbPE}{[11]}{}, but we expect non-champions may require more than generic stories: They need something to disrupt and capture their attention, something that differs from common well-known media stories they already dismissed as irrelevant to their work. We  aim to create stories that not only illustrate harms in the moment but also resonate deeply and are remembered later to seed an actual transformation, not just a short-term boost in engagement. 

To achieve such impact, we turned to marketing theory, grounding our approach in the principles of \emph{Made to Stick} \gencite{JO4x}{[31]}{}, which identifies characteristics of memorable and persuasive ideas. We operationalized these principles into five key story qualities to guide the construction and evaluation of our \textbf{sticky stories}:

\begin{compactitem}
	\item \textbf{\surprise \ Surprisingness.} Stories should present harms in ways that disrupt default assumptions, that are counterintuitive or non-obvious. This quality captures attention, which is a prerequisite for reflection and potential attitude change. Evidence from cognitive science shows that unexpected stimuli attract attention and trigger deeper processing \gencite{yxsG}{[35]}{}. By making harms surprising, practitioners are more likely to notice risks that would otherwise be overlooked and experience disorienting dilemmas.

	\item \textbf{\concrete \ Concreteness.} Stories should include tangible details, such as named roles, real-world analogues, or observable system behaviors. Concreteness helps audiences visualize harms and form emotional connections, making abstract principles more understandable and memorable. Dual Coding Theory \gencite{tt74}{[20]}{} supports this approach, showing that concrete information is encoded both verbally and visually, improving recall compared to abstract descriptions.

	\item \textbf{\severe \ Severity.} Severity emphasizes the magnitude and scope of potential harm, highlighting why certain outcomes demand attention and action. Perceived severity motivates engagement, as individuals are more likely to respond to risks they judge serious. Research on the affect heuristic \gencite{HRDr}{[85]}{} demonstrates that people's risk judgments are strongly influenced by the emotional weight of outcomes, with severe consequences eliciting stronger reactions and prompting protective or corrective behaviors. 

	\item \textbf{\relevant \ Relevance.} Stories should align with domain-specific experiences and stakeholder concerns, ensuring that messages feel applicable to practitioners' work. Relevance increases the likelihood that examples are processed deeply and influence attitudes or behaviors \gencite{zhbR}{[50]}{}. By situating lessons in authentic professional contexts, practitioners can connect the scenarios to real decisions, enhancing engagement and reflection.

	\item \textbf{\diverse \ Diversity.} A broad range of stakeholders, harm types, and system behaviors can avoid narrow or stereotypical portrayals. Evidence from narrative transportation research indicates that encountering multiple perspectives enhances engagement and supports attitude change \gencite{3tPw}{[29]}{}. Diverse stories help practitioners anticipate harms across contexts rather than focusing on isolated examples.

\end{compactitem}

\paragraph{Capturing Early Signs of Change: Critical Reflection.}
 Our intervention used \emph{sticky stories} to create \emph{disorienting dilemmas}, aiming to prompt reflective thinking. While our ultimate goal is \emph{transformation}, genuine perspective shifts are difficult to observe without extended observation windows over multiple years; immediate responses may not indicate lasting change. To keep the scope of our research manageable, we focus on \emph{early indicators of transformation}, observing moments when \emph{disorienting dilemmas} triggered \emph{critical reflection}.

To observe signs of critical reflection, beyond just short-term measures of engagement, we focused on concrete behavioral signs that participants were moving beyond surface-level reactions. Prior work defines \emph{critical reflection} as moving beyond descriptive or casual reflection to actively examine one's assumptions, beliefs, and actions, evaluating their validity and potential consequences \gencite{pdsS}{[53]}{}. In the context of responsible AI, this would involve questioning default practices, recognizing ethical risks, and considering how one's work may contribute to harm. To systematically detect these moments of \emph{critical reflection}, we identified \emph{concrete behavioral indicators,} summarized in Table~\ref{tab:criticalreflection}.  We will use these indicators in our evaluation in Sec. 6, tracing how these moments manifested during participants' engagement with the stories.

\tableReflection{Challenges assumptions \gencite{pdsS}{[53]}{}}{Explores multiple perspectives \gencite{zD9z}{[10]}{}}{Connects to wider systems or past incidents \gencite{zD9z}{[10]}{}}{Expresses surprise \gencite{pdsS}{[53]}{}}{Engages in iterative thinking \gencite{L12W}{[83]}{}}{Plans intentional change \gencite{8xcg}{[52]}{}}

\subsection{Generating Sticky Stories }\label{h.eltx9xxp614a}
\figureRAI

Generating high-quality \emph{sticky stories} is non-trivial. We found that single zero-shot or few-shot prompts with LLMs were not very effective in generating stories that meet all five of our criteria, as they often produce outputs that are overly generic, repeat common tropes (e.g., race and gender only), and fail to capture surprising or contextually relevant harms. Without structured guidance, LLMs struggle to systematically combine diverse harm types, non-obvious stakeholders, and generate stories that effectively provoke reflection. This motivated our design of a systematic and scalable compound AI system \gencite{1Io5}{[90]}{}  that combined prompt engineering techniques with programmatic control to ensure that each story consistently embodied the five desired qualities. We broke down the task into a coherent sequence of logical steps and iterated through multiple design cycles to refine each component. This resulted in the following eight-step pipeline (cf.~Fig.\ref{fig:rai}):

\begin{compactitem}
	\item \textbf{Input (\relevant).} To make the stories relevant to non-champions, we grounded them in the specific projects participants were working on. As inputs, we collected the ML system's description, its intended purpose, and a representative stakeholder use case. These inputs seeded the rest of the pipeline, ensuring that the generated stories were \emph{relevant}---directly tied to the participant's own ML system rather than abstract or generic examples.

	\item \textbf{Step 1: Pre-define Harm Types (\diverse).} To ensure diversity and coverage of edge cases, we began by specifying a set of harm categories. Pre-defining these categories grounds story generation in well-theorized frameworks of harm rather than ad hoc examples. For this, we drew on the taxonomy of fairness-related harms from prior studies of harm categories \gencite{vbPE,WtFk}{[11, 84]}{} and fairness goals from Microsoft's RAI assessment guide \gencite{jLLo}{[55]}{}. The set included cultural misrepresentation, reinforcement of biases, unequal access to opportunities, and erasure of minorities.

	\item \textbf{Step 2: Identify Stakeholders (\surprise, \diverse).} To generate stories that capture diverse harms, it is also essential to identify stakeholders whom practitioners might otherwise overlook. For example, in a movie recommendation system, practitioners often focus on obvious stakeholders such as movie watchers, but may miss stakeholders such as movie producers, whose livelihoods depend on whether their films are surfaced. To account for these cases, we go beyond conventional direct and indirect stakeholders by introducing \emph{direct-surprising} and \emph{indirect-surprising} categories. These highlight marginalized or non-obvious groups, following the principles of \emph{Design Justice} \gencite{vKL0}{[14]}{}, which emphasizes attention to those often overlooked in design processes. We prompt an LLM to generate these different sets of stakeholders based on the product description. By surfacing unexpected stakeholders, the stories are more likely to create \emph{disorienting dilemmas} that challenge practitioners' default assumptions.

	\item \textbf{Step 3: Identify Demographics (\concrete, \relevant).} To ground stories in concrete,  contexts, relevant to the user, we then generate possible demographic attributes for each stakeholder (e.g., age, gender, ethnicity). This step facilitates subsequent steps in tailoring harms to marginalized or contextually relevant groups.

	\item \textbf{Step 4: Generate Initial Stories (\diverse).} To increase diversity in our stories, for each harm--stakeholder combination, we generate an initial set of harm stories, forming a matrix of harms and affected users. This matrix-based approach, inspired by prior work \gencite{vbPE}{[11]}{}, ensures broad coverage and combinatorial richness. By systematically exploring combinations, we reduce the risk of narrow or stereotypical examples, and instead highlight harms that may not surface through ad-hoc brainstorming. 

	\item \textbf{Step 5: Regenerate for Surprisingness (\surprise).} Given that the LLM might default to producing stories that align with patterns it frequently encounters, to avoid bland or generic outputs, we prompt the model to regenerate stories using earlier outputs as counterexamples---encouraging less typical, more striking, and surprising narratives.

	\item \textbf{Step 6: Cluster and Sample for Diversity (\diverse).} To further increase diversity and avoid redundancy, we transform the stories into sentence embeddings, apply K-means clustering (k=10), and sample from the five least-populated clusters---those most likely to contain unique narratives.

	\item \textbf{Step 7: Refine for Concreteness and Severity (\concrete, \severe).} Concreteness makes harms tangible and easier to visualize, increasing practitioners' emotional engagement. To make sure the stories are concrete and severe, we employ a two-stage refinement loop: one model refines stories for concreteness and severity, and a second evaluates the output. Stories lacking specificity or clarity are iteratively revised (up to three times) to meet our concreteness standard.

	\item \textbf{Step 8: Pick Stories with the Highest Severity (\severe).} Rather than enumerate every harm comprehensively (as other tools \gencite{vbPE,73Qw}{[11, 97]}{} and hazard analysis \gencite{FdBE,msWG}{[42, 79]}{} pursue), we aim to provoke and persuade with a small number of high-impact stories. Therefore, in a final step, we prompt an LLM to select the two stories with the greatest magnitude and scope of harm, while ensuring they satisfy all five qualities. 

\end{compactitem}

\subsection{Sticky Story Integration in a Tool }\label{h.9ohsla7k0lag}
To demonstrate how sticky stories could be presented to practitioners and to run our evaluation study, we integrated the pipeline into an interactive tool. The tool is designed to replicate Microsoft's RAI assessment guide \gencite{jLLo}{[55]}{} (Fig.~\ref{fig:template}), which is typically completed as a static, text-based template.

Users begin by entering a brief description of the ML system, its intended purpose, a user story, and system stakeholders (Fig.~\ref{fig:tool}-A (1, 2)). Subsequently, during the fairness assessment step (Fig.~\ref{fig:tool}-B and C), users can request brainstorming assistance, which presents the previously generated sticky stories (Fig.~\ref{fig:tool}-D) for the current assessment step.

To reduce potential delays and maintain a smooth user experience, these stories are often generated in earlier screens of the tool---based on the user's description of the ML system---and stored for retrieval in the subsequent fairness brainstorming step. This approach can help ensure that stories appear quickly when requested (story generation with the GPT-4o model typically takes 3--4 minutes) helping users focus on the task without waiting---though in practice the timing may vary depending on system load and context. The tool shows two stories by default, but users can request up to three more stories, provide feedback, and regenerate stories.

\figureTemplate{\emph{Snapshot of Microsoft's Responsible AI assessment template \gencite{jLLo}{[55]}{}}}

\figureTool

\section{Evaluation I: Evaluating Stickiness of Harm Stories}\label{h.1w7ijon1n3mb}
We first conducted an offline evaluation to understand the quality and cost of generating sticky stories with our designed pipeline. In the evaluation, we curate diverse AI application scenarios and run our pipeline and an ablated version to generate the harm stories. We then measure the quality of the generated stories with the five desired qualities of sticky stories: \textbf{concrete (\concrete), severe (\severe), surprising (\surprise), diverse (\diverse), and relevant (\relevant),} as well as the cost of generating these stories in terms of token usage and time elapsed.

\subsection{Experiment Setups}\label{h.p7wt1ytbxwc9}
\subsubsection{Data}\label{h.ni4571baw6qt}
We collected diverse AI application scenarios from the Internet (e.g.,  \gencite{VZQi,iJzE,MwFl}{[57, 68, 69]}{}) and randomly sampled 15 scenarios (e.g., \emph{voice assistants}, \emph{image search}, \emph{email monitoring}, and \emph{demand forecasting}) for our evaluation. 

We used an LLM (\texttt{gpt-4o}) to process the searched content into more detailed descriptions, similar to what a user would have input to our system, and we manually verified that these generated descriptions are valid. 

\subsubsection{Methods}\label{h.wqivdh2rw0k3}
For evaluating the generation capabilities, we implement most of our pipeline with \texttt{gpt-4o}, as it demonstrates strong writing capabilities \gencite{EeFD}{[63]}{}. In step 7, however, we use a smaller model \texttt{gpt-4o-mini} as the evaluator to reduce cost, as validation usually requires less capabilities than generation. We use \texttt{mxbai-embed-large-v1} to produce sentence embeddings for K-means clustering.

\emph{Baseline.} We compare our pipeline approach to a zero-shot prompting baseline, in line with prior work \gencite{vbPE}{[11]}{}. The prompt directly instructs an LLM (\texttt{gpt-4o}) to generate scenarios that illustrate harm to relevant stakeholders, and we instruct that the baseline prompts generate stories around 175 words, which is the average length we observe from the stories generated by our pipeline. We share all prompts used in our supplementary material.

\emph{Story generation.} For each AI application scenario, we generate two stories for each of the two fairness goals (\emph{Quality of Service}, and \emph{Allocation of Resources and Opportunities}). In total, we curated 120 sticky stories and 120 baseline stories. This sample size allows us to draw conclusions with 90\% confidence level with 8\% margin of error.

\subsubsection{Metrics}\label{h.pyq32u7g6m4m}
We evaluate the quality of the sticky stories and baseline stories and the cost of the generation method. For cost, we measure the time it takes to run the pipeline and the number of tokens it costs. For quality, we evaluate the stories on the five desired qualities of \emph{sticky stories}. Four quality metrics (\emph{concrete}, \emph{severe}, \emph{surprising}, \emph{relevant}) are evaluated with a two-phase evaluation involving both human raters and LLMs, with diversity evaluated by a separate established distance metric. The final evaluation covered \textbf{240 stories} produced by both methods.

\paragraph{Human Annotation and Inter-Rater Reliability.}
 Three researchers independently evaluated a set of 20 harm stories using binary (yes/no) judgments across the four qualities of stickiness. We conducted multiple calibration rounds, during which we refined the definitions of the qualities and clarified edge cases based on observed disagreements. After we reached consensus, we developed the finalized rubric (see supplementary material) for a larger-scale evaluation.

\paragraph{Scalable Evaluation Using LLM-as-a-Judge.}
 To scale the evaluation to the full set of generated stories, we adapted our rubric for an automated evaluation using \texttt{gpt-4o}. The prompt included the story text, plain-language definitions of each of the qualities, and a binary decision task for each dimension (0 = no, 1 = yes). Four of the five dimensions were rated using binary LLM judgments. 

To further validate the reliability of the LLM judgments, one researcher independently annotated 30 additional stories. The Cohen's Kappa values for agreement between human and GPT-4 annotations were: Concrete: 1.000, Severity: 0.4783, Surprising: 0.9356, and Relevant: 0.8387. All disagreements on Severity stemmed from the LLM being overly generous in rating baseline stories as severe, this overinflating the severity results of the baseline (see Figure~ref{fig:matrices}). 

\figureMatrices

\paragraph{Measuring Diversity with Embedding Distances.}
 For diversity, we computed cosine distances between semantic embeddings of the story titles (higher distance indicates higher diversity), as it is a well-established distance measurement in the literature \gencite{9FgN}{[72]}{}. 

\subsubsection{Limitations (Threats to Validity)}\label{h.7vufcmo8i02k}
Despite extensive validation, internal validity may be affected by biases in LLM-based judgements. Our binary ratings for each quality captures only big differences and may not represent more nuanced quality differences. External validity is constrained by the small, curated set of 15 scenarios, which may not represent all AI applications. 

\subsection{Results}\label{h.20b5h3igviu3}
\tableOfflineEval

Overall, we found that our pipeline is able to generate sticky stories that are more concrete (+98.3\%), more severe (+30.9\%), and more surprising (+42.9\%) than baseline stories, demonstrating the effectiveness of our design (see Table~\ref{tab:offlineeval} for more details). The sticky stories are also generally more diverse (+5.8\%), due to our clustering approach. However, we do observe a small trade-off in relevance (-8.7\%), as sometimes the generated sticky stories are overly dramatic and can be hard to relate. In addition, generating sticky stories requires more resources (5.5x time and 46x token usage) than the zero-shot generation of baseline stories. As we will show next, this cost is likely acceptable, as the sticky stories are indeed more effective  at engaging practitioners and inspiring them to think of RAI harms beyond their existing mindsets.

\section{Evaluation II: User Study}\label{h.uipcsf2hv2hb}
After showing that sticky stories indeed embody the five desired qualities (Sec. 5), we now assess their practical value, that is, whether sticky stories actually lead to greater engagement from (non-champion) practitioners. We conducted a user study that explored how non-champion practitioners engage with harm identification tasks across three conditions: without support, with baseline stories, and with \emph{sticky stories}. We analyzed differences in terms of (1) the time participants spent identifying harms, (2) the number of new harm categories they surfaced, and (3) the depth and nature of their \emph{critical reflections} on the harms or stories.

\subsection{Study Design}\label{h.w8hpx5hg1hst}
To evaluate the impact of \emph{sticky stories} on practitioner engagement, we conducted a \textbf{mixed-design user study} that combined both \textbf{within-subject} and \textbf{between-subject} elements. Unlike prior work that often focuses only on champions, we deliberately sought to include non-champions, and ended up with a range of practitioners with varied levels of RAI motivation. Each participant completed two harm identification tasks under two of three conditions: \emph{no stories}, \emph{baseline stories}, or \emph{sticky stories}. This design allowed us to disentangle the effect of story presence from the unique qualities of sticky stories, while also partially controlling for potential learning and ordering effects. We assessed engagement through multiple indicators, including time spent, number and diversity of harms surfaced, and qualitative markers of reflection in response to the stories.

\subsubsection{Participants and Recruitment}\label{h.d1hot9oi6qb6}
Unlike prior studies that did not account for self-selection bias, which likely led to primarily recruiting participants already motivated by RAI concerns, our evaluation required recruiting \emph{non-champion} practitioners---those less inclined to prioritize RAI in their work. Recruiting this group was inherently difficult, as they are unlikely to volunteer for a study framed around Responsible AI. To overcome this, we strategically oversampled through broad advertisements that emphasized ML evaluation broadly, and probed participants' preferences for different evaluation techniques in a screening survey. This design choice helped attract a broader and more neutral practitioner audience, including individuals who are less inclined toward fairness assessments and thus more representative of non-champions whom our intervention seeks to engage. Recruitment and study protocols were approved by our Institutional Review Board (IRB).

We recruited participants through professional platforms such as LinkedIn, Twitter, and a large Slack community for data scientists. Interested individuals completed the screening survey, which included questions regarding their familiarity with concepts such as model training, model evaluation, model fairness, AI ethics, and MLOps, and how useful they think various evaluation activities are, including in-distribution data evaluation, out-of-distribution data evaluation, model red-teaming, and responsible AI auditing (e.g., fairness). This enabled us to identify practitioners who do not prioritize RAI in their work. We received a total of 291 responses. We excluded submissions that indicated low engagement or fraudulent behavior---such as vague project descriptions, suspicious email addresses, or missing LinkedIn profiles---and filtered participants based on their stated level of high RAI interest. We conducted 5 pilot experiments to test and refine the study protocol. Afterward, we recruited 31 participants, but due to a tool malfunction in which the baseline stories failed to generate, data from two participants could not be analyzed. That is, we successfully conducted the study with \textbf{29 participants}. Each participant received as compensation a \$35 gift card.

\subsubsection{Experimental Conditions}\label{h.wi0qqao3fdc4}
To evaluate the impact of the sticky story intervention on practitioner engagement, we designed three experimental conditions. Engagement could potentially be influenced either by the presence of any illustrative story or specifically by the \emph{sticky stories}. To disentangle these effects, we implemented two control conditions and one treatment condition.

\begin{compactitem}
	\item \textbf{Condition A (No Stories - Control 1):} Participants complete a section of the RAI assessment without being shown any stories.

	\item \textbf{Condition B (Baseline Stories - Control 2):} Participants complete a section of the RAI assessment while being shown two baseline harm stories (see Sec. 5.1.2; zero-shot prompting, matched for length of sticky stories).

	\item \textbf{Condition C (Sticky Stories - Intervention):} Participants complete a section of the RAI assessment while being shown two \emph{sticky stories}.

\end{compactitem}

\subsubsection{Tasks}\label{h.9lrrg1ikezvu}
Each participant completed two tasks focused on harm identification and mitigation planning. To ensure ecological validity \gencite{68OO}{[38]}{} and help participants engage more deeply with the task that is realistic and personally relevant, participants analyze \emph{their own projects.} Each task asks the participant to identify harms and possible mitigations for one of two fairness goals from Microsoft's RAI assessment guide \gencite{jLLo}{[55]}{}---a well-established framework developed independently of our study:

\begin{compactitem}
	\item \textbf{Fairness Goal 1:} Allocation of resources and opportunities.

	\item \textbf{Fairness Goal 2:} Minimization of stereotyping, demeaning, and erasing outputs.

\end{compactitem}

Each participant worked on both tasks in a random order. Analyzing their own projects ensures that the evaluation reflects realistic, personally relevant contexts.

\subsubsection{Treatment Groups}\label{h.lhhvwghl89z}
\figureStudy

We randomly assigned participants to one of four treatment groups. Each group began with a task in the no story condition (Condition A), followed by the other task in either the baseline (Condition B) or sticky story condition (Condition C),  see Fig~\ref{fig:study}. This design enables both \textbf{within-subject} comparison (e.g., engagement with vs. without stories) and \textbf{between-subject} comparison (baseline vs. sticky stories). In addition, the time where participants worked on the task without stories would allow us to precompute stories for the other task. By randomizing assignment and counterbalancing the order of fairness goals, we reduce confounds related to task complexity and learning effects.

\subsubsection{Study Protocol}\label{h.lnurxo1lplnj}
Each participant completed a pre-study survey, answered a few questions to establish their background, worked on the two tasks with and without stories, and finally debriefed with the facilitator. This usually took about 60 minutes. Two months after the experiment, we sent a follow up survey.

\paragraph{Pre-Study: Participant Background and RAI Orientation.}
 We collected background information to understand each participant's experience and relationship with RAI practices (e.g., their exposure to RAI), both through a short survey and a brief verbal discussion (see \emph{Interview Guide} in supplementary material). Drawing on  stated choice research \gencite{LQOf}{[44]}{}, we ask questions about behaviors rather than ask about preferences, focusing on  \emph{revealed preferences} (from their actual practices, past behaviors, and decisions regarding fairness and harm mitigation) rather than \emph{stated preferences} (participants' expressed values, concerns, and attitudes toward RAI).

\paragraph{During the Study: Think-Aloud Harm Identification.}
 While the participants worked on the two tasks, we employed a \textbf{think-aloud protocol} to capture participants' real-time thought processes and reasoning while engaging with harm identification tasks, allowing us to understand not just what they identify but how they interpret and respond to different stories. We issued only minimal prompts to preserve the validity of participants' cognitive processes during harm identification tasks \gencite{iie5,hXR0}{[8, 23]}{}. The facilitator maintained a neutral stance and refrained from influencing participants' reasoning---intervening only when they misunderstood the task or deviated significantly from it. In conditions where participants were exposed to stories (either \emph{baseline} or \emph{sticky}), we deliberately avoided leading questions to minimize bias. Instead, we used minimal, open-ended prompts (e.g., ``\emph{What are your thoughts as you read this}?'' or ``\emph{Feel free to keep thinking aloud}'') to encourage continued verbalization and self-reflection. 

\paragraph{Post-Study: Reflections and Intentions.}
 Immediately after completing the tasks, we invited participants to reflect on their experience by answering open-ended questions about how the task influenced their understanding of fairness-related harms and whether they intended to take any concrete actions or revisit decisions in their own projects. These reflections allowed us to capture participants' intentions to act---serving as a proxy for how compelling, relevant, or actionable they found the experience.

\paragraph{Post-Study Follow-Up (2 Months Later).}
 To assess whether the intervention led to any sustained engagement or reflection, we conducted a follow-up survey two months after the main study. In the survey, participants were asked two open-ended questions:

\begin{compactitem}
	\item \emph{Have you reacted to or done anything based on the findings from our session (e.g., changed anything in your past or new projects directly or indirectly based on concerns raised during our discussion)?}

	\item \emph{Have you had any discussions---positive or negative---about responsible AI with your peers since the session?}

\end{compactitem}

While insufficient to measure long-term transformations (which may require years), this still captures some concrete \textbf{behavioral and social impact over time}, even if modest, to understand whether the participants engaged with RAI concepts in their professional communities beyond our study.

\subsection{Data Analysis}\label{h.20nk7o1oivi6}
To evaluate how practitioners engage differently across study conditions, we defined specific goals and aligned our data sources accordingly. Our primary goal was to understand engagement and reasoning in response to baseline versus sticky stories. We operationalized this goal using targeted questions paired with corresponding metrics (Table~\ref{tab:gqm}), following the established Goal-Question-Metric (GQM) approach \gencite{paty}{[4]}{}. We combined quantitative metrics (e.g., harm diversity, time spent, user characteristics) with qualitative coding to characterize participants' engagement and reasoning processes (see the finalized variables in Table~\ref{tab:variables}).

\tableGQM

\tableVariable{\gencite{vbPE}{[11]}{}}{Table~\ref{tab:criticalreflection}}

\paragraph{Quantitative Analysis.}
 To evaluate whether practitioners invest more time when exposed to sticky versus baseline stories---a common proxy for engagement, e.g., \gencite{koOw}{[26]}{}---we extracted the total time spent in each section along with the free-text harms from tool logs. To measure harm diversity, each harm was manually assigned to a category and subcategory from an established RAI taxonomy \gencite{vbPE}{[11]}{}. We assigned each participant a score regarding their prior RAI awareness and championship, based on self-reported experience in the pre-study survey and our assessment of their answers to our initial questions (revealed preferences), which we used as controls in our analysis. Table~\ref{tab:variables} summarizes the variables used in our study. We use ANOVA to analyze the influence of the experimental condition on the dependent variables. Model diagnostics---including checks for normality, homogeneity of variance, and influential points---were performed to ensure the validity of the analyses.

\paragraph{Qualitative Analysis.}
 To understand \emph{how} participants engaged with harm identification tasks---beyond what quantitative metrics could capture---we conducted \emph{qualitative content analysis} \gencite{KL77,n7Vr}{[33, 41]}{}, a method that affords quantitative analysis of qualitatively coded data. We used a carefully designed codebook combining theory-driven indicators and patterns emerging from the transcripts (Table~\ref{tab:criticalreflection}). In particular, deep engagement was coded using the indicators of critical reflection described in Section 4.1, supplemented with additional codes for shallow engagement. Two independent coders applied the codebook to the think-aloud transcripts, and inter-rater reliability was calculated to refine the codes, yielding a Cohen's \(\kappa\) of 0.72, which indicates substantial agreement. Codes were then applied systematically with each participant's transcript as a chunk of analysis, that is,  multiple mentions by the same participant were not counted repeatedly. Follow-up survey responses were also examined to capture participants' self-reported takeaways and any indications---planned or already undertaken---of applying these insights in their own projects.

\subsection{Limitations (Threats to Validity/Credibility)}\label{h.4yi0sv2yobs}
As every study, ours needs to make tradeoffs and has limitations. Our study captures short‑term engagement rather than lasting behavior change; even with a two‑month follow‑up, we cannot establish long‑term effects. The modest sample constrains statistical power, and may limit generalizability across domains and organizations. Biases are possible at several levels: social desirability (despite neutral prompts, avoidance of ``like/dislike'' questions, and focusing on revealed preferences), bias in participant selection (even with broad, neutral framing, oversampling, screening for non‑champions, and limited snowballing), and researcher bias in qualitative interpretation (despite standard research practices, codebook, dual coding, and reporting inter‑rater reliability). Internally, the mixed design places the no‑story control first for all participants, so any story is confounded with appearing second; learning, priming, fatigue, and carryover from the first fairness goal may influence observed gains. The think‑aloud protocol can also alter problem‑solving strategies and make engagement appear higher than in ordinary work. Our engagement proxies---time on task and counts of distinct harm categories---are imperfect; time can reflect confusion, and breadth can trade off with depth. Finally, while anchoring tasks in participants' own projects improves realism, it introduces heterogeneity that can mask or mimic effects, and the study examines a single researcher‑facilitated exposure rather than repeated use in everyday workflows. We deliberately designed the study, trading off various qualities and accepting some limitations; readers should interpret our results accordingly.

\subsection{Findings}\label{h.rwy9pvolut8o}
\subsubsection{Finding 1: Practitioners spent significantly more time on harm identification tasks when sticky stories were shown}\label{h.313sx9mjiajo}
\tableDescriptive

\tableAnovaTime

Participants with \emph{sticky stories} for their second task spent substantially more time on harm identification tasks compared to those with \emph{baseline} \emph{stories}, with very large effects observed. Participants with baseline stories increased the time spent modestly over their task in the \emph{no-story condition} (+11\%, from 7.5 to 8.3 minutes), whereas participants with \emph{sticky stories} spent nearly triple the time compared to their first task in the \emph{no-stories condition} (+207\%, from 5.4 to 16.6 minutes); see Table~\ref{tab:descriptive}, Fig.~\ref{fig:timeChangeChart}.

Statistical analyses controlling for task order, experience, awareness, and championship confirm a very large effect of story type on the relative time increase between the first (no story) and second task (see Table~\ref{tab:anovatime}). An additional analysis suggests that the time participants spend on the first task influences the time they spend on the second task (i.e., some participants are generally slower/more thorough than others), but story type accounted for substantially more variance, confirming that sticky stories associated clearly with more time spent on the second task.

\figuretimeChangeChart

While we anticipated some increase in time for sticky story participants, the magnitude of the increase even over baseline stories was surprising. Qualitative examination of tool logs, think-aloud transcripts, and video recordings revealed distinct patterns of engagement. Participants with baseline stories generally skimmed the stories, often moving quickly through the task without much deliberation. In contrast, participants with sticky stories typically processed each story sequentially, critically evaluating its applicability. Many paused to reflect on past incidents, consider stakeholder perspectives, or connect the story to broader systemic issues, indicating sustained cognitive engagement rather than superficial completion. We discuss this further in \emph{Finding 4 and 5}. 

\subsubsection{Finding 2: Practitioners identified significantly more diverse harms when sticky stories were shown}\label{h.5lbgwc7m67ox}
\tableAnovaHarm

Participants generally listed small numbers of harms as result from each task (typically 0 to 5 harms, cf. Table~\ref{tab:descriptive}, Fig.~\ref{fig:histogram}). While we see an increase in the number of harms identified when sticky stories were provided (but not for baseline stories), we do not find counting the number of harms itself very meaningful, as participants might repeat very similar harms.  Instead, we focused on the diversity of harms---measured as the number of distinct harm categories.

Participants exposed to \emph{sticky stories} identified substantially more harm categories and subcategories than those in the \emph{baseline} \emph{stories} \emph{condition}, corresponding to roughly 4.5× more new categories and 3.5× more new subcategories (see Table~\ref{tab:descriptive}, Fig.~\ref{fig:violin}). These differences remained statistically significant after controlling for task order, awareness, prior championship, and experience in an ANOVA, suggesting that the observed increases were robust across participant backgrounds and prior familiarity with fairness concepts.

\figureHistogram

\figureViolin

To understand the reasons behind the differences, we analyzed participants' responses before and after exposure to the stories, alongside their recorded interactions with them. Participants in the \emph{sticky story} \emph{condition} were more likely to surface new harms and expand across categories: of the six participants (\PFour, \PFifteen, \PSeventeen, \PTwentyOne, \PTwentyFive, \PThirty) who left the harms section blank in the task 1 (\emph{no-story condition)}, all but the \emph{baseline} participant, added harms in task 2. Several also reconsidered fairness goals they had initially dismissed. For instance, \POne began by rejecting the goal of minimizing stereotyping---``To be honest, I don't think this applies to this system''---but after reading a story, exclaimed, \emph{``Oh, man, this is topical! My wife's [country] [...] this could be true},'' and added stereotyping-related harms. All participants who expanded into multiple new harm categories (\POne, \PTwo, \PFour, \PThirteen, \PSeventeen, \PTwentyOne, \PTwentyThree, \PThirty) were in the \emph{sticky story condition}. In some cases, \emph{sticky stories} also helped participants who were otherwise ``stuck'' in a single harm category---for instance, \PFive initially listed four harms exclusively under \emph{Allocational Harms}, but in task 2, added one under Quality of Service Harms. By contrast, participants in the \emph{baseline condition} often disengaged in task 2, leaving the section blank (\PTwentyFive, \PTwentySeven) or typing perfunctory responses (e.g., \PTwentyTwo: ``It is as was mentioned in the generated scenarios''). Even many who recorded harms (\PThree, \PTen, \PFourteen, \PNineteen, \PTwenty, \PTwentyTwo, \PTwentyFive, \PTwentySeven, \PTwentyEight, \PThirtyOne) largely repeated categories they had already noted, showing little expansion of perspective.

\subsubsection{Finding 3: Practitioners critically reflected on the sticky stories more, but only skimmed the baseline stories}\label{h.me86cmv9aiar}
Based on the distribution of critical reflection codes, participants in the sticky story group engaged more deeply and reflected more often than those in the baseline group (Fig.~\ref{fig:codes}). Among the indicators of critical reflection (Table~\ref{tab:criticalreflection}), the most common code was \emph{expressing surprise or enthusiasm}, observed in 17 participants overall---15 of whom were from the \emph{sticky story condition}. Participants expressed surprise with remarks such as \emph{``oh, wow''} (\PFifteen, \PNineteen), \emph{``Oh, my God, I mean, that could have caused an issue''} (\PNine), or \emph{``I didn't really think about the fairness or responsible AI aspect of my system… I was really surprised''} (\PThirtyOne).

The second most frequent code was \emph{connecting to the wider system or past incidents}, which also occurred more often in the \emph{sticky story group} (9 participants) than in the baseline group (4 participants). Participants made these connections either by recalling past incidents---e.g., \emph{``If I deploy [the biased model] then [...] it's gonna cost millions of dollars for internal systems. It happened some time back [...] [company] got shut down [...] millions of dollars of impact''} (\PTwentyThree) or anticipating wider implications.

Notably, behaviors such as \emph{challenging assumptions (}e.g., \PTwentySix: \emph{``Now that I'm looking at the word demeaning [...] there's definitely some of this happening}''), \emph{exploring multiple perspectives (}e.g., \PFifteen: ``\emph{how can a government or the FDA respond?}''\emph{)}, and \emph{iterative thinking} appeared exclusively in the \emph{sticky story group}. 

In contrast, signs of shallow engagement---such as skimming or dismissing the stories, or attempting to copy them directly as harms without further reasoning---were only observed in the baseline condition (e.g., \PSeven, \PTen, \PTwentyFive, \PTwentySeven, \PTwentyEight, \PTwentyNine). These participants often spent minimal time with the scenarios and focused on completing the task quickly (consistent with \emph{Finding 1}). 

Six participants across both groups voiced some negative reactions to at least one story. However, the tone and follow-up differed markedly. Baseline participants tended to be dismissive---e.g., \emph{``I find them very general''} (\PTwentySeven)---whereas sticky story participants more often contextualized or justified their critiques. For instance, \PTwentySix initially found one story \emph{``a little far-fetched''} but immediately connected it to their own experience of harassment, and \PFive expressed doubt but then qualified it with technical reasoning about their system's design.

\figureCodes

\subsubsection{Finding 4: Follow-up survey revealed more post-study actions than the sticky story group}\label{h.nzmjtz64fjug}
In total, we sent follow-up emails to 25 of the 29 participants; the remaining four had not yet passed the two-month gap period at the time of paper submission. Nine participants responded---six from the sticky story group (\POne, \PTwo, \PNine, \PSeventeen, \PNineteen, \PTwentyFour) and three from the baseline group (\PThree, \PFourteen, \PTwentyTwo). While these responses are not sufficient to support statistical conclusions, we report them here for completeness and transparency.

Responses from the sticky story group consistently described concrete changes to practice, particularly in data collection and quality processes. These included recruiting more diverse participants for data collection (\emph{P1}), conducting thorough safety checks before adding data for fine-tuning or training models using LLM-as-a-judge \gencite{nfjC}{[103]}{} (\PSeventeen), enhancing data quality pipelines with expert review (\PNineteen), and adopting stricter labeling standards and rigorous annotation quality checks (\PTwentyFour). Several also highlighted measures to improve transparency and mitigate harm, such as adding explanations and feedback mechanisms to reduce biased outcome (\PTwo), implementing multiple layers of safety and legal filters (\PSeventeen), and adopting a more deliberate approach to risk assessment (\PNine). \PNine further noted that their team had updated its Terms \& Conditions to explicitly outline safeguards for stakeholders. Many also reported ongoing discussions with colleagues about fairness, harm mitigation, and ethical risks---for example, \PTwentyFour shared, \emph{``Yes, I have had several discussions on responsible AI with my research team and colleagues.''}

In contrast, baseline group responses were fewer and generally less action-oriented. One participant described a vague intention to ``\emph{use AI more responsibly}'' without a concrete plan (\PThree), another admitted to making no changes, and a third (\PFourteen) mentioned future plans to integrate responsible AI considerations but had not yet acted on them, though they had shared key insights from the study with their organization's responsible AI team.

\subsubsection{Finding 5: Practitioners exhibit distinct trajectories in shifting from initial indifference or resistance toward a more engaged stance on RAI}\label{h.e0u84je73qap}
By examining participants' behaviors and narratives, we posit five distinct engagement profiles:

\begin{compactitem}
	\item \resistor \ \emph{Resistors}: Explicitly push back against RAI as irrelevant or hype (\PSeventeen, \PThirty).

	\item \indifferent \ \emph{Indifferents}: Acknowledge RAI is important but show little follow-through (\PTwo, \PThree, \PFive, \PSix, \PTwentyOne, \PTwentySix).

	\item \follower \ \emph{Followers}: Follow RAI practices as their company enforces it (\POne, \PFourteen, \PNineteen, \PTwentyThree, \PTwentyFour, \PTwentyFive, \PTwentySeven).

	\item \learner \ \emph{Learners}: Have limited prior knowledge of RAI, but eager to learn (\PSeven, \PEight, \PNine, \PTen, \PFifteen, \PTwenty, \PTwentyTwo, \PTwentyEight, \PThirtyOne).

	\item \champion \ \emph{Champions}: Already motivated and advocate of RAI (\PFour, \PEleven, \PTwelve, \PThirteen, \PTwentyNine)

\end{compactitem}

\paragraph{\resistor \ Resistors.}
 We identified only two participants (\PSeventeen, \PThirty) who were explicitly dismissive of RAI, describing it as hype and irrelevant to their work. This small number was not surprising, given the likelihood of \emph{social desirability bias}---many who held similar sentiments may have instead presented themselves as \indifferent \ \emph{Indifferents}. Both resistors were in the sticky story group, meaning we cannot draw conclusions about how baseline participants in this category might have behaved. Notably, however, both individuals demonstrated a notable trajectory: 

\trajectory{Outright dismissal}{Skepticism}{Recognition of overlooked risks}{Reframing as RAI-relevant issues (expanded awareness)}

This trajectory illustrates how \emph{sticky stories} can move even resistant practitioners toward expanded awareness, with \emph{skepticism} emerging as the stories create \emph{disorienting dilemmas} that challenge participants' existing views. 

To illustrate this trajectory, we highlight the case of \PSeventeen, who initially rejected RAI as unrelated to their domain, saying, ``\emph{I don't really believe in it, to be honest. In my area of research, the societal harms are very low. […] I'll give the cliché example---the prison system predicting recidivism across races. But that has nothing to do with code generation.''} This perspective carried into the \emph{no-story condition}, where they spent only two minutes and recorded no harm. The introduction of \emph{sticky stories} shifted this pattern. Although initially skeptical---\emph{``It's hypothetically possible, but very implausible''}---they subsequently acknowledged overlooked risks, such as bias toward non-English codebases: \emph{``Sure, you can overfit to English. This is actually true.''} By the end, \PSeventeen conceded that what they had considered ``just tool issues'' could fall under RAI: \emph{``I didn't know that certain things fall under RAI.''} This shows how sticky stories can provoke expanded awareness even among practitioners who begin with strong resistance.

\paragraph{\indifferent \ Indifferents.}
 Indifferents are participants whose stated preference in responsible AI (RAI) contrasted sharply with their revealed preference (i.e., they rate it as important in a survey, but their described past behaviors do not suggest active engagement). We speculate that they may acknowledge RAI's importance mostly in a general sense or due to social desirability. Within this profile, we observed distinct patterns between the \emph{sticky story} and \emph{baseline} \emph{groups}. Participants in the \emph{baseline group} (\PThree, \PSix) demonstrated minimal, superficial engagement, whereas those exposed to \emph{sticky stories} (\PTwo, \PFive, \PTwentyOne, \PTwentySix) engaged more deeply, reflected on the scenarios, and generated concrete, actionable plans, following the trajectory: 

\trajectory{Indifference}{Curiosity sparked}{Critical reflection}{Planned concrete actions}

This curiosity may again stem from \emph{disorienting dilemmas}, triggered by \emph{diverse} (\diverse) and \emph{surprising} (\surprise) cases they had not anticipated.

To illustrate this trajectory, we highlight the case of \PFive. Initially, without stories, \PFive spent 7 minutes on the first task, engaging mechanically and offering broad, vague assessments of potential harms, such as unreliable model performance or general stakeholder impacts. Their proposed mitigations were equally generic, such as improving system accuracy. After exposure to \emph{sticky stories},  \PFive's engagement deepened substantially, spending 31 minutes carefully working through five stories (two initial and three additional ones). They revisited each story multiple times, expressed surprise and curiosity, and began considering a wider range of potential risks and systemic consequences: \emph{``I hadn't considered how stakeholder decisions would change based on these predictions.''} Their reflections evolved from general concerns to concrete, nuanced mitigation strategies, such as supporting developing regions with specialist guidance.  Through this process, \PFive transitioned from indifference to critical reflection, ultimately generating actionable plans they could realistically implement, showing how \emph{sticky stories} can spark genuine engagement and meaningful planning even among initially indifferent participants.

In contrast, \PThree, a representative participant in the baseline condition in this profile, showed little evidence of meaningful engagement. They spent 3 minutes on the \emph{no-stories condition} and 5 minutes with the \emph{baseline stories}, and the harms they identified before and after were nearly identical, suggesting minimal reflection. Rather than using the stories as prompts for deeper consideration, they approached them mechanically, even remarking, \emph{``I would say, scenario 3 is the most relevant scenario. So can I just copy and paste it.''} This shallow interaction might be due to  the limits of generic story prompts in stimulating reflection, especially compared to the richer engagement observed with sticky stories.

\paragraph{\follower \ Followers.}
 These practitioners operate in organizations with established RAI infrastructures---mandatory harm assessment, governance teams, and formal review processes. Their engagement with RAI was often shaped more by institutional requirements than personal motivation, which often made the processes feel bureaucratic and burdensome. In the baseline group (\PFourteen, \PTwentyFive, \PTwentySeven), this compliance-driven stance generally persisted, with participants identifying only surface-level harms and routine mitigations (e.g., \PTwentyFive: ``Okay, is this the end of it?''). By contrast, we saw a noticeable shift in participants in the sticky story group (\POne, \PNineteen, \PTwentyThree, \PTwentyFour): The stories encouraged them to move beyond ``checking the boxes'' of oversight toward deeper reflection on risks, systemic consequences, and their own role in addressing them (e.g., \PTwentyThree: \emph{``I wasn't expecting it to go this deep… I really liked it.'')}. For some, this shift translated into a new sense of ownership and proactive responsibility, as they began connecting stories to gaps in their work and articulating concrete actions they intended to take. This reflects a trajectory similar to that of the Indifferents: 

\trajectory{Compliance mindset}{Assumptions challenged / Recognition of stake}{Critical reflection}{Proactive responsibility}

We conjecture that \emph{severity} (\severe), \emph{relevance} (\relevant), and \emph{concreteness} (\concrete) of harm stories are particularly important to move participants beyond their routine view of RAI.

\paragraph{\learner \ Learners. }
Participants in this profile entered with only a vague awareness of Responsible AI. Many had heard of fairness concerns in the abstract, but lacked concrete exposure to what such issues actually look like in practice. Within this profile, we saw two orientations: (a) some participants did not initially recognize the gaps in their knowledge or the need to deepen it (\PSeven, \PEight, \PFifteen, \PThirtyOne), while (b) others openly admitted limited familiarity but expressed curiosity and willingness to explore (\PNine, \PTen, \PTwenty, \PTwentyTwo, \PTwentyEight). Although the sample size is too small for statistical claims, anecdotally, participants in the sticky story group showed a strikingly larger learning shift compared to those in the baseline group (e.g., \PFifteen: ``\emph{I didn't know this could happen. It's also new for me. But being a data scientist, that's not good.'')}. Their engagement suggested the following trajectory: 

\trajectory{Unawareness}{Recognition of gaps (group a) / Broadening of perspective  (group b)}{Proactive involvement}{}

This mirrors the trajectory from unconscious incompetence to conscious incompetence in pegagogy literature \gencite{ovNp}{[25]}{}. We suspect  \emph{surprisingness} (\surprise) and \emph{diversity} (\diverse) of stories is particularly important here.

\paragraph{\champion \ \emph{Champions}.}
 While our study primarily aimed to recruit non-champions, we classified some of our participants as  champions when we talked to them (\PFour, \PEleven, \PTwelve, \PThirteen, \PTwentyNine). These individuals did not inherit RAI responsibilities through a formal role but still were intrinsically motivated to pursue RAI, often driving initiatives without organizational incentives. Among this group, we did not observe substantial differences in harm identification across conditions. One participant (\PTwentyNine) only skimmed the baseline stories before moving directly to identifying harms independently, but all others engaged with the stories (baseline or sticky) and critically reflected on them. Although the small sample size prevents drawing statistical conclusions, these observations match our expectations: champions are already motivated to engage deeply with RAI assessments, whereas non-champions (\resistor, \indifferent, \follower, \learner) seem to need a little push to spark meaningful reflection and action.

\section{Discussion}\label{h.ohztewect7ti}
Our findings suggest that sticky stories can meaningfully influence RAI engagement, significantly more so than simple zero-shot generated examples of harms. In this section, we discuss who benefits, how they benefit, and why this matters.

\paragraph{Who Benefits: Differential Impact on Champions and Non-Champions.}
 The few champions (\champion) in our study were already highly engaged with RAI tasks, so exposure to sticky stories had little impact on their harm identification or reflection. In contrast, non-champions (\resistor, \indifferent, \follower, \learner) showed noticeable changes when interacting with sticky stories, with increases in time spent, harms identified, and critical reflection. This suggests that while existing interventions may be effective among already motivated individuals, non-champions benefit from additional prompts to trigger \emph{disorienting dilemmas}---and sticky stories appear to provide that push.

\paragraph{How They Benefit: Impact of Story Qualities Across Practitioner Profiles.}
 Moreover, the observed trajectories suggest that different story qualities may resonate differently with non-champion profiles. \emph{Learners} (\learner) may benefit most from diverse (\diverse) stories that expose them to new possibilities; \emph{followers} (\follower) may be more influenced by severe (\severe) stories that highlight potential consequences; \emph{indifferents} (\indifferent) may respond primarily to surprising (\surprise) stories that challenge expectations; and \emph{resistors} (\resistor) may require a combination of surprising (\surprise) and relevant (\relevant) stories that create both dissonance and relevance to shift their engagement. We do not imply that any single quality is sufficient---transformation may depend on multiple qualities acting together---but certain qualities may have stronger effects depending on practitioner type. This points to an important direction for future work: systematically evaluating how individual story qualities affect different practitioner profiles.

\paragraph{Why It Matters: Potential for Long-Term Transformation.}
 Transformation, according to Transformative Learning Theory \gencite{pdsS}{[53]}{}, is a multi-stage long journey beginning with a (a) \emph{disorienting dilemma} that (b) \emph{challenges prior assumptions}, followed by (c) \emph{self-examination} and (d) \emph{critical assessment} of those assumptions, (e) \emph{recognition of shared experiences} with others, (f) \emph{exploration of new roles}, (g) \emph{planning a course of action}, (h) \emph{trying out new roles}, (i) \emph{building competence} in those roles, and (j) finally \emph{reintegration} of these changes into one's perspective and behavior. Research also suggests that sustaining mindset change typically requires repeated reinforcement, ongoing organizational support, and integration into routine practices \gencite{8Ote,RJM7}{[66, 98]}{}. In this study, although we designed sticky stories with longer-term transformation in mind, we could only demonstrate short-term effectiveness and have a few encouraging signs of mid-term benefits from our two-months follow up. Our evaluation primarily captured immediate engagement within a single session, leaving open the question of whether effects persist once the novelty fades. While observed trajectories hint at initiation of several stages of Transformative Learning (e.g., a--f, and some g), these observations alone are insufficient to confirm durable transformation. Nonetheless, these findings are encouraging, suggesting that sticky stories could be particularly effective if incorporated directly into practitioners' regular harm assessment workflows---such as embedding them within checklists or model cards---so they serve as recurring prompts rather than one-off interventions. In this way, sticky stories can continue to reinforce awareness, stimulate peer discussion, and normalize reflection on responsible AI, bridging the gap between immediate engagement and sustained behavioral change.

\end{document}